\documentclass[prb,superscriptaddress,longbibliography,twocolumn]{revtex4-1}
\usepackage{amsmath}
\usepackage{amssymb}
\usepackage{graphicx}
\usepackage{hyperref}
\usepackage{braket}
\usepackage{bm}
\usepackage{color}

\newcommand{\eqsa}[1]{\begin{eqnarray} #1 \end{eqnarray}}

\begin{document}

\title{Classical and quantum spin dynamics of the honeycomb $\Gamma$
  model}

\author{Anjana M. Samarakoon}
\affiliation{Neutron Scattering Division, Oak Ridge National Laboratory, Oak Ridge, Tennessee 37831, U.S.A.}
\affiliation{Shull-Wollan Center, Oak Ridge National Laboratory, Oak Ridge, Tennessee 37831, U.S.A.}
\author{Gideon Wachtel}
\affiliation{Department of Physics, University of Toronto, Toronto,
  Ontario M5S 1A7, Canada}
\affiliation{Racah Institute of Physics, The Hebrew University, Jerusalem 91904, Israel}
\author{Youhei Yamaji}
\affiliation{Department of Applied Physics, The University of Tokyo, Hongo, Bunkyo-ku, Tokyo, 113-8656, Japan}
\affiliation{JST PRESTO, Hongo, Bunkyo-ku, Tokyo, 113-8656, Japan}
\author{D.~A.~Tennant}
\affiliation{Materials Science and Technology Division, Oak Ridge National Laboratory, Oak Ridge, Tennessee 37831, U.S.A.}
\affiliation{Shull-Wollan Center, Oak Ridge National Laboratory, Oak Ridge, Tennessee 37831, U.S.A.}
\author{Cristian D. Batista}
\affiliation{Neutron Scattering Division, Oak Ridge National Laboratory, Oak Ridge, Tennessee 37831, U.S.A.}
\affiliation{Shull-Wollan Center, Oak Ridge National Laboratory, Oak Ridge, Tennessee 37831, U.S.A.}
\affiliation{Department of Physics and Astronomy, University of Tennessee, Knoxville, Tennessee 37996-1200, U.S.A.}
\author{Yong Baek Kim}
\affiliation{Department of Physics, University of Toronto, Toronto,
  Ontario M5S 1A7, Canada}
\affiliation{Canadian Institute for Advanced Research, Quantum Materials Program, Toronto, Ontario M5G 1M1, Canada}
\affiliation{School of Physics, Korea Institute for Advanced Study, Seoul 130-722, Korea}

\begin{abstract} 
  Quantum to classical crossover is a fundamental question in dynamics
  of quantum many-body systems. In frustrated magnets, for example, it
  is highly non-trivial to describe the crossover from the classical
  spin liquid with a macroscopically-degenerate ground-state manifold,
  to the quantum spin liquid phase with fractionalized
  excitations. This is an important issue as we often encounter the
  demand for a sharp distinction between the classical and quantum spin
  liquid behaviors in real materials. Here we take the example of the
  classical spin liquid in a frustrated magnet with novel
  bond-dependent interactions to investigate the classical dynamics,
  and critically compare it with quantum dynamics in the same
  system. In particular, we focus on signatures in the dynamical spin
  structure factor. Combining Landau-Lifshitz dynamics simulations and
  the analytical Martin-Siggia-Rose (MSR) approach, we show that the
  low energy spectra are described by relaxational dynamics and highly
  constrained by the zero mode structure of the underlying degenerate
  classical manifold. Further, the higher energy spectra can be
  explained by precessional dynamics. Surprisingly, many of these
  features can also be seen in the dynamical structure factor in the
  quantum model studied by finite-temperature exact
  diagonalization. We discuss the implications of these results, and
  their connection to recent experiments on frustrated magnets with
  strong spin-orbit coupling.
\end{abstract}

\maketitle

\section{Introduction}

The crossover between classical and quantum regimes in frustrated
magnets has been an important theoretical question in the last few
decades. This issue is particularly important in understanding the
nature of the quantum spin liquid phases which may arise at low
temperature due to extreme quantum fluctuations. In the classical
regime, there may exist a window of temperatures below the Curie-Weiss
scale, where the correlated spin moments are thermally fluctuating
within the degenerate manifold of classical ground states.  This is
the cooperative paramagnetic state, or the so-called classical spin
liquid.  In the quantum regime, it is clearly not possible to maintain
such a state down to zero temperature as the quantum ground state
should be unique (up to a topological degeneracy in the case of
quantum spin liquids).  An important question is how much information
about the degenerate manifold of classical ground states is encoded in
the emergent quantum spin liquid at zero and low temperatures.  Such a
question may be especially relevant for two dimensional spin liquid
phases which show no finite temperature phase transition, but only
crossovers.

One natural place to look for the clue for this question is the
dynamical spin correlation or the dynamical spin structure factor. A
recent work on the Kitaev model\cite{kitaev_anyons_2006} in two
dimensions investigates the dynamical spin correlations of the
classical Kitaev model via Landau-Lifshitz (LL)
dynamics~\cite{Samarakoon17}. The resulting dynamical structure factor
was compared with that of the quantum model\cite{knolle_dynamics_2014,
knolle_dynamics_2015, baskaran_exact_2007}, which is exactly solvable
and supports a quantum spin liquid ground state with gapless Majorana
fermion excitations.  There exist two crossover temperatures, $T_v$
and $T_Q$, in the Kitaev model on the honeycomb lattice, as seen in
the specific heat.\cite{PhysRevB.92.115122} At $T < T_v$, the vison or
flux gap is larger than the temperature scale so that the system is
essentially characterized by the zero temperature ground state. When
$T_v < T < T_Q$, the flux degree of freedom is thermally disordered,
but the Majorana fermions are still well defined. For $T > T_Q$, the
system crosses over to the classical regime. It was found that the
dynamical spin correlations in the quantum model at $T > T_v$ are
remarkably similar to those of the cooperative paramagnetic regime of
the classical model at finite temperature.  Moreover, the dynamical
structure factor of the quantum model knows about the zero mode
structure of the classically degenerate manifold even when $T _v < T <
T_Q$.  This suggests that all the classically degenerate spin states
are participating in the quantum fluctuations down to $T \sim T_v$,
which eventually lead to the emergence of the quantum spin liquid
phase at low temperature $T < T_v$.

In principle, it is not necessary that the full degenerate manifold of
the classical states is involved in the formation of the quantum spin
liquid at low temperature, since thermal entropy or zero-point quantum
fluctuations may select a subset of the full degenerate manifold at
some intermediate temperature.  On the other hand, when the full
degenerate manifold is participating in quantum fluctuations at low
temperature, as the case of the Kitaev model for $T_v < T < T_Q$, and
if the spin correlations remain short-ranged, it is highly suggestive
that the zero temperature ground state would indeed be a quantum spin
liquid.  An alternative choice for the zero temperature ground state
could be a magnetically ordered state or a quantum critical point,
which would show the development of long-range dynamical spin
fluctuations. In the case of the Kitaev model, we already know that
this is not the case, and that the zero temperature ground state is a
quantum spin liquid. One may, however, be able to use this lesson to
infer the possible presence of a quantum spin liquid in models which
are not exactly solvable.

  In the current work, we investigate the dynamical spin correlations
in the classical and quantum $\Gamma$ model\cite{rau_generic_2014}
(defined below) on the honeycomb lattice, which is known to possess
macroscopically degenerate manifold of classical ground
states\cite{rousochatzakis_classical_2017}, while the quantum model is
not exactly solvable. This model represents the bond-dependent
anisotropic and symmetric spin interactions on the honeycomb lattice:
  \begin{equation} {\cal H} = \;\; \mathrm{\Gamma }\sum_{\alpha \neq
\beta \neq \gamma }{\sum_{\left\langle {\bm r}, {\bm r}' \right\rangle
\ \in \ \gamma }{\mathrm{(}S^{\beta }_{\bm r} S^{\alpha }_{{\bm
r}'}\mathrm{+}S^{\alpha }_{\bm r} S^{\beta }_{{\bm r}'} \mathrm{)}}},
    \label{gammamodel}
  \end{equation} where $S^\alpha_{\bm r}$ are spin operators at sites
$\bm{r}$ of a honeycomb lattice, and $\alpha,\beta,\gamma=x,y,z$. Such
interactions (as well as the Kitaev interaction mentioned above) arise
in strongly spin-orbit coupled Mott insulators such as Li$_2$IrO$_3$
and $\alpha$-RuCl$_3$, where Ir$^{4+}$ or Ru$^{3+}$ ions form
effective $J=1/2$ local moments.  Currently, the relative importance
of the Kitaev and $\Gamma$ interactions is an important issue in
theoretical and experimental investigations of this class of
Kitaev-like materials\cite{jackeli_mott_2009,ybkim_2014,
rau_generic_2014, plumb_2014, sears_magnetic_2015,
johnson_monoclinic_2015, rau_spin-orbit_2016, winter_challenges_2016,
winter_models_2017, trebst_2017, sears_phase_2017,
do_incarnation_2017, hentrich_unusual_2018, wolter_field-induced_2017,
winter_breakdown_2017, winter_probing_2018}.  For instance, the 
dominance of one of these interactions or the cooperation of these
two interactions may lead to possible emergence of quantum spin liquid
in these materials, especially in the presence of external pressure or
magnetic field.

We first use the Landau-Lifshitz dynamics to compute the dynamical
structure factor in the classical model at finite temperature.  It is
shown that the zero mode structure of the degenerate manifold of
classical ground states is reflected in the low frequency part of the
dynamical spin fluctuation spectra.  For example, in the case of the
antiferro-sign of the $\Gamma > 0$ interaction, the structure factor
at low frequencies is suppressed at the $\bm \Gamma$ and $\bm X$
points of the Brillouin zone, which we explain using the constraints
on the classical spin states which belong to the degenerate manifold.
Next, we employ the Martin-Siggia-Rose (MSR) formulation of Langevin dynamics
to further understand the nature of the dynamical spin correlations.
These two different methods lead to essentially the same dynamical
spin correlations, leading to the conclusion that the system itself
may be acting as an effective thermal bath. Furthermore, it is shown
that the low frequency response is relaxational and reflects the zero
mode structure, while the higher frequency response is mostly
precessional, and some characteristic precessional modes exist at
finite frequencies.  The evolution of dynamical spin correlations is
also investigated as a function of temperature 
for the comparison with the quantum model.

The dynamical spin structure factor in the quantum $\Gamma$ model is
studied by exact diagonalization via the shifted Krylov subspace
method, which is combined with typical quantum state approach at
finite temperature. In previous
studies\cite{catuneanu_path_2018}, it was shown that there exist
two crossover temperatures, $T_1 \sim 0.03 \Gamma$ and $T_2 \sim 0.4
\Gamma$, in the specific heat, similar to the case of the Kitaev
model.  $T_2$ marks the crossover from the high temperature classical
regime to the quantum regime.  We find that the dynamical spin
correlations at low frequencies in the quantum model show distinct
signatures of the zero mode structure of the degenerate manifold of
classical ground states even when $T_1 < T < T_2$, which gradually
crosses over to the low temperature extreme quantum limit for $T <
T_1$. This behavior is reminiscent of the dynamical spin correlations
in the Kitaev model, where correlations as a function of temperature
are remarkably similar to the classical result.  This means that the
short-range spin fluctuations from the degenerate manifold persist
even in the quantum regime of $T_1 < T < T_2$.  On the other hand, the
dynamical spin correlations at zero temperature, while they remain
short-ranged, show features that are not present in the classical
model.  In the case of $\Gamma$ model, we currently do not know what
the quantum ground state is. One possibility is that below a certain
temperature a symmetry is broken due to order by quantum
disorder\cite{rousochatzakis_classical_2017}. Interestingly, however,
recent DMRG and exact diagonalization studies suggest possible
presence of a quantum spin liquid ground state in the $\Gamma$
model\cite{gohlke_quantum_2018}.  Since we do not have an analytical
understanding of the underlying quasiparticles in the $\Gamma$ model,
we cannot make a more precise connection to underlying quantum degrees
of freedom, which was possible in the case of the Kitaev model.
Nonetheless, the phenomenological similarity to the classical-quantum
correspondence in the Kitaev model is striking.  We may speculate that
the entire degenerate manifold of classical states would participate
in quantum fluctuations that lead to the formation of the quantum
ground state at zero temperature, such as a quantum spin liquid
phase. Our findings will help understanding this outstanding issue and
possible connection to experiments on Kitaev-like materials.

The rest of the paper is organized as follows. In section
\ref{sec:LL}, we present numerical results obtained from the LL
dynamics of the classical model. Section \ref{sec:lan} describes how
qualitatively similar results can be obtained analytically within an
MSR formalism. The dynamic correlations of the corresponding quantum
model are given in section \ref{sec:ed}. We conclude with a discussion
of our main findings in light of recent experiments in section
\ref{sec:dis}, while details of our calculations are relegated to the
appendices.

\section{Landau-Lifshitz dynamics}
\label{sec:LL}

We study the dynamical spin correlations of the $\Gamma$ model,
Eq. (\ref{gammamodel}),
whereby the spin operators are replaced by classical vectors in the
Heisenberg equation of motion. The resulting Landau-Lifshitz (LL)
equation of motion is
\begin{equation} \frac{d {\bm S}_{\bm r}}{dt}\mathrm{=}{\bm S}_{\bm r}
\mathrm{\times }{\bm B}_{\bm r},
\label{eq:LL}
\end{equation} where ${\bm B}_{\bm r}$ is the molecular field 
acting on the spin ${\bm S}_{\bm r}$. This LL equation can be solved
numerically by applying a fourth order Runge-Kutta algorithm with
adaptive step size. The average over configurations at a given
temperature $T$ is obtained from the Metropolis Monte Carlo sampling
method. We note that a closed LL dynamics is more appropriate than
Langevin dynamics when the experiment is much faster than the
spin-lattice relaxation, which is the typical case in inelastic
neutron-scattering experiments. The simulations are performed on a
finite lattice of 30$\times$30 unit-cells (1800 spins) with periodic
boundary conditions.

\begin{figure}[htp]
  \centering
  \includegraphics[width=0.45\textwidth]{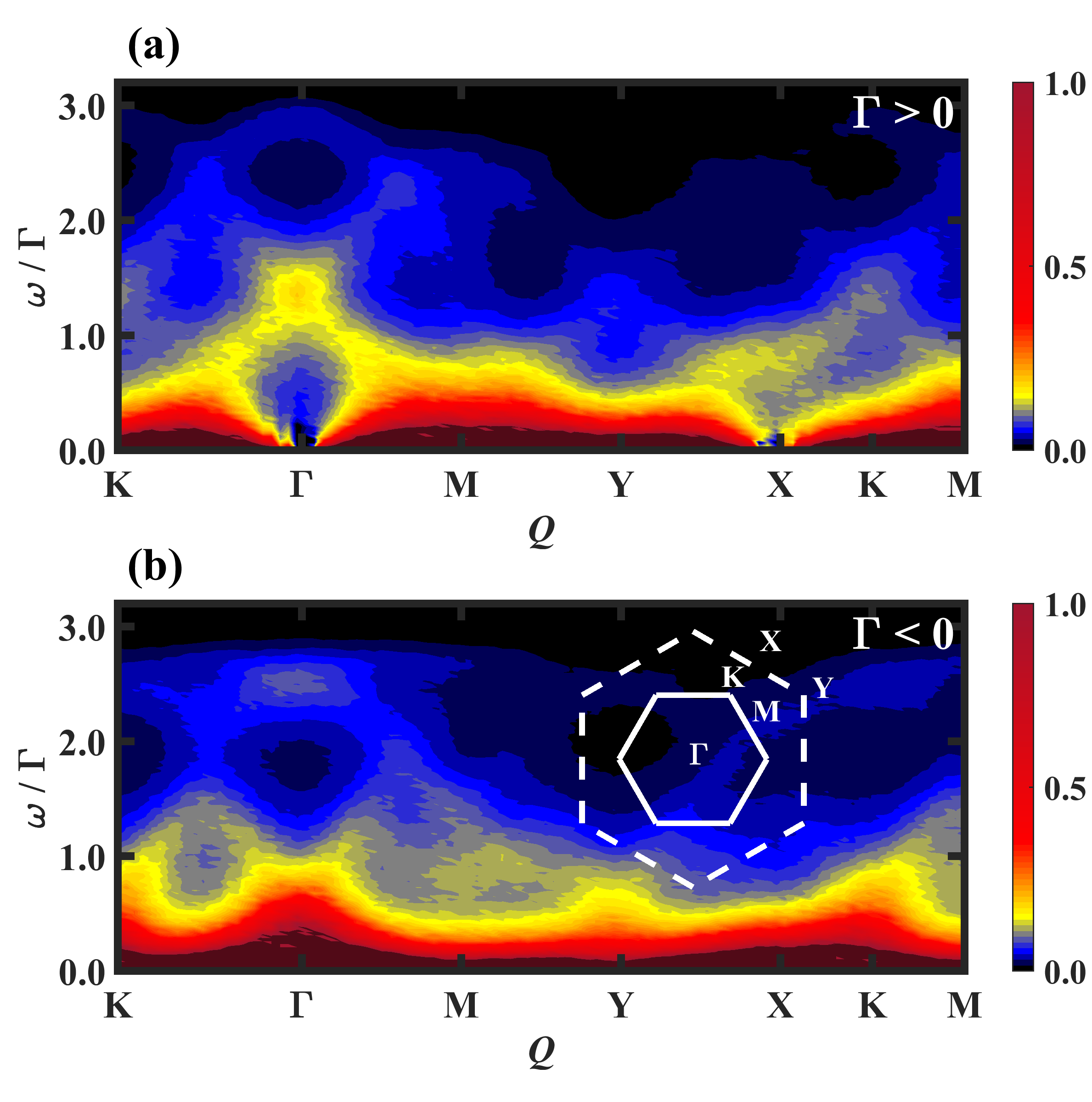}
  \caption{Trace of the dynamical magnetic structure factor
    $S\left({\bm Q},\omega \right)=\ \sum_{\alpha }{S^{\alpha \alpha
      }\left({\bm Q},\omega \right)}$ for (a) AFM ($\mathrm{\Gamma
    }>0$) (b) FM ($\mathrm{\Gamma }<0$) $\Gamma$-models along the
    Brillouin Zone path, ${\bm K}-{\bm \Gamma}-{\bm M}-{\bm Y}-{\bm
      X}-{\bm K}-{\bm M}$, as depicted in the inset.}
  \label{fig01}
\end{figure}

Fig.~\ref{fig01} shows the trace of the dynamical spin structure
factor, $S\left( {\bm Q},\omega \right)=\ \sum_{\alpha }{S^{\alpha
\alpha }\left({\bm Q},\omega \right)}$, obtained for the
antiferromagnetic (AFM) (${ \mathrm \Gamma }>0$) and ferromagnetic
(FM) ($\mathrm{\Gamma }<0$) versions of the $\Gamma$-model.  As
expected for a liquid state, $S\left( {\bm Q},\omega \right)$ exhibits
a continuum of low and high-frequency modes.  The low-frequency (zero)
modes arise from the very slow dynamics through different classical
ground states. The number of zero modes is macroscopic because of the
extensive residual entropy of the ground state manifold. This dynamics
is expected to be {\it relaxational} because the average of the local
field over a period $2 \pi/\omega$ is equal to zero. 
In contrast, the high-frequency modes correspond to the much faster
spin precession around the local fields produced by {\it a given
  ground state configuration}. 
Accordingly, the average of the local fields ${\bm B}_{\bm r}$ over a
period $2 \pi/\omega$ remains finite.

\begin{figure}[htp]
  \centering
  \includegraphics[angle=0,clip,width=1.0\columnwidth]{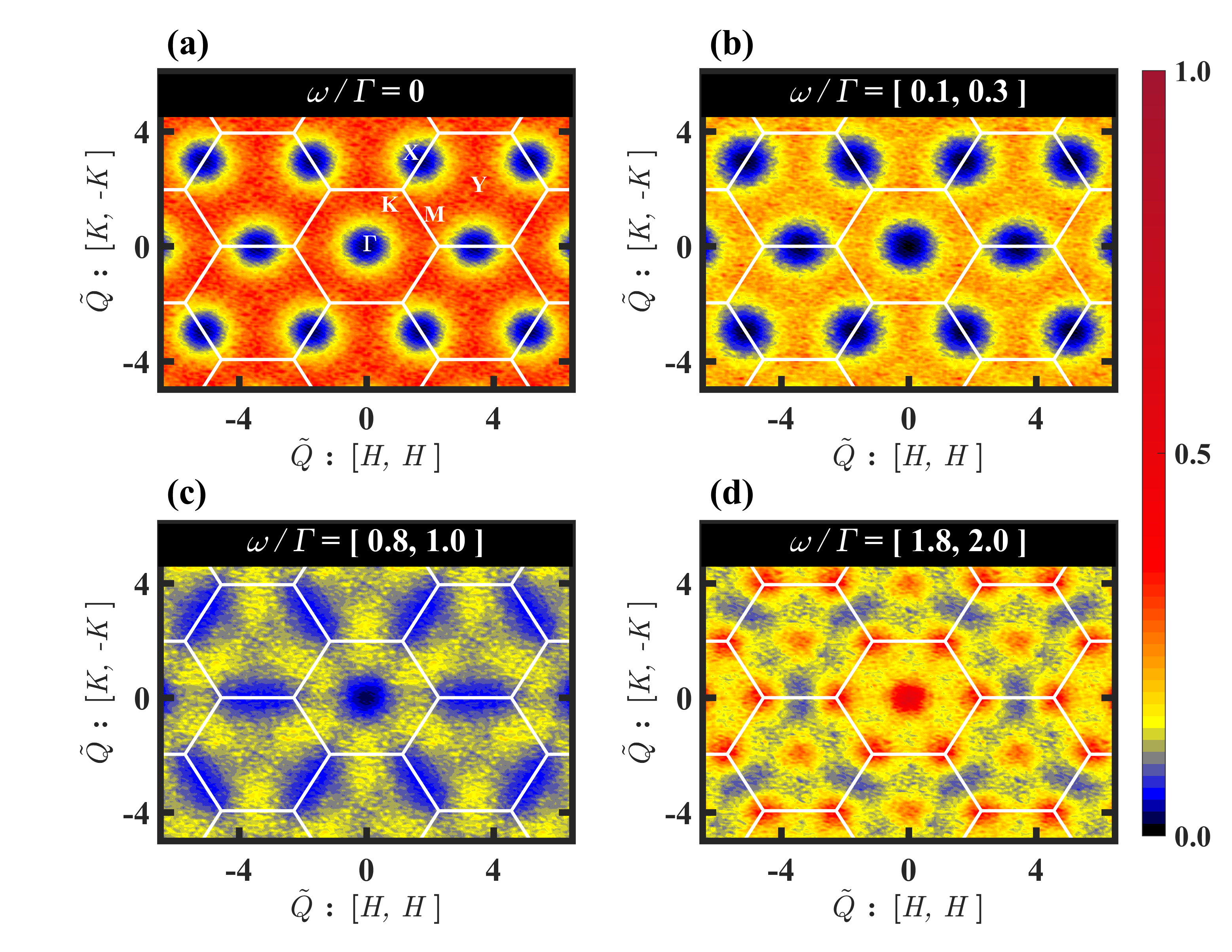}
  \caption{(a) Elastic component of the trace of the magnetic
    structure factor $S({\bm Q}, \omega=0)$. (b-d) 
    $S({\bm Q},\omega )$, integrated over finite energy cuts: (b)
    $\omega /\mathrm{\Gamma } = [0.1, 0.3]$, (c) $\omega
    /\mathrm{\Gamma }$ = [0.8, 1.0] and (d)$\ \omega /\mathrm{\Gamma
    }$ = [1.8, 2.0].}
  \label{fig02}
\end{figure}

Both, the low and high-frequency modes, contain relevant information
about the liquid state. The momentum distribution of the zero-energy
modes is a direct consequence of the set of constraints defining the
ground state manifold.  Specifically, we show in Appendix
\ref{sec:zeromodes} that the fourier transform $S^\alpha({\bf
  q})=\sum_{\bf r}S_{\bf r}^\alpha e^{i{\bf q}\cdot{\bf r}}$ of any
state $S^\alpha_{\bf r}$ in the ground state manifold, vanishes for
both ${\bf q} = \bm\Gamma$ and ${\bf q} = \bf{X}$.
As a result, the low-energy spectral weight of $S^{\alpha \alpha } (
{\bm Q},\omega)$ is suppressed at the ${\boldsymbol \Gamma}$ and ${\bm
  X}$ points of the Brillouin zone [see
Fig.~\ref{fig02}~(a)]. Correspondingly, as shown in
Fig.~\ref{fig02}~(b)-(d), the missing low-energy spectral weight at
these two points is shifted to frequencies of order $\Gamma$. In other
words, magnetic excitations with wave vectors ${\boldsymbol \Gamma}$
and ${\bm X}$ are purely precessional. Indeed, as shown in
Fig.~\ref{fig02}, the precessional modes have highest intensity at
these two wave vectors.  As we will see later, the low-energy modes of
the quantum $S=1/2$ model inherit this structure.  The high-energy
modes contain information about the magnitude and spatial distribution
of the instantaneous local fields ${\bm B}_{\bm r}$ of a typical
ground state spin configuration. The dispersion of these modes
contains information about the magnetic correlation length of the spin
liquid state.

\begin{figure}[htp]
  \centering
  \includegraphics[angle=0,width=1.0\columnwidth]{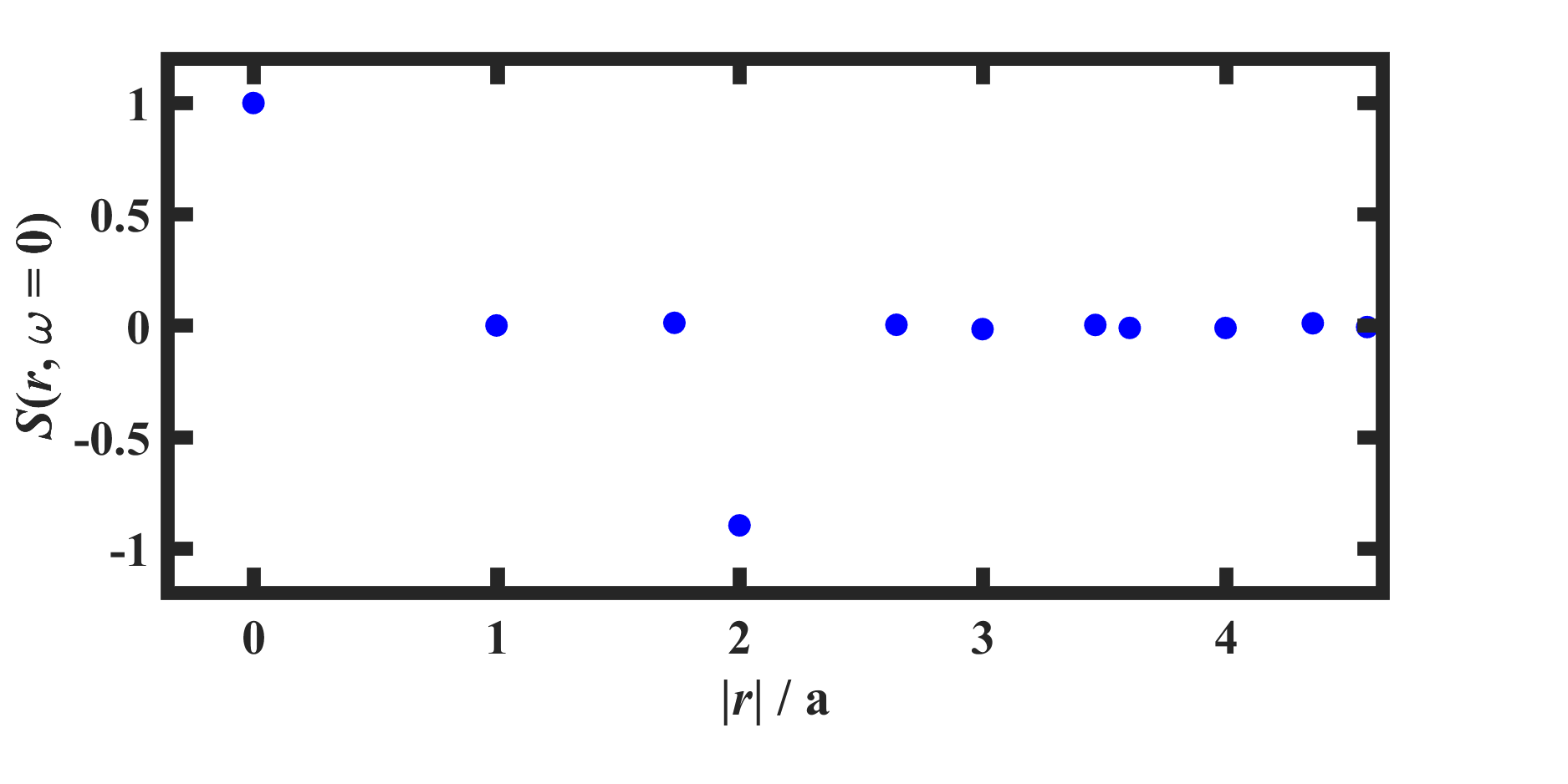}
  \caption{ $S({\bm r},\omega=0 )$ as a function of ${|{\bm
        r}|/a}$ (where $a$ is the lattice parameter) at $T= 10^{-5}
    \Gamma$.
  }
  \label{fig03}
\end{figure}

To gain more insight on the structure of the zero-energy modes, 
we also present the real space spin-spin correlation function, $S({\bm
  r},\omega )$, as a function of $ \omega $ and $T$.  Fig.~\ref{fig03}
shows the elastic contribution $S({\bm r},\omega=0 )$ for different
distances up to fifth nearest neighbors (NN) and $T= 10^{-5} \Gamma$
(calculations need to be done at finite $T$ to have fluctuations and
be able to exploit the fluctuation-dissipation theorem).  Remarkably,
this is significant only for the on-site and for the
third-nearest-neighbor ($|{\bm r}|/a=2$, where $a$ is the lattice
parameter) correlation functions.  The values obtained for other
distances are smaller than the statistical error of the MC
calculation. Therefore, the fast exponential decay of the elastic
contribution indicates that the magnetic correlation length is of
order $2a$.

\begin{figure*}[htp]
  \centering
  \includegraphics[trim={0cm 0cm 0cm 0cm},clip,width=1.0\textwidth]{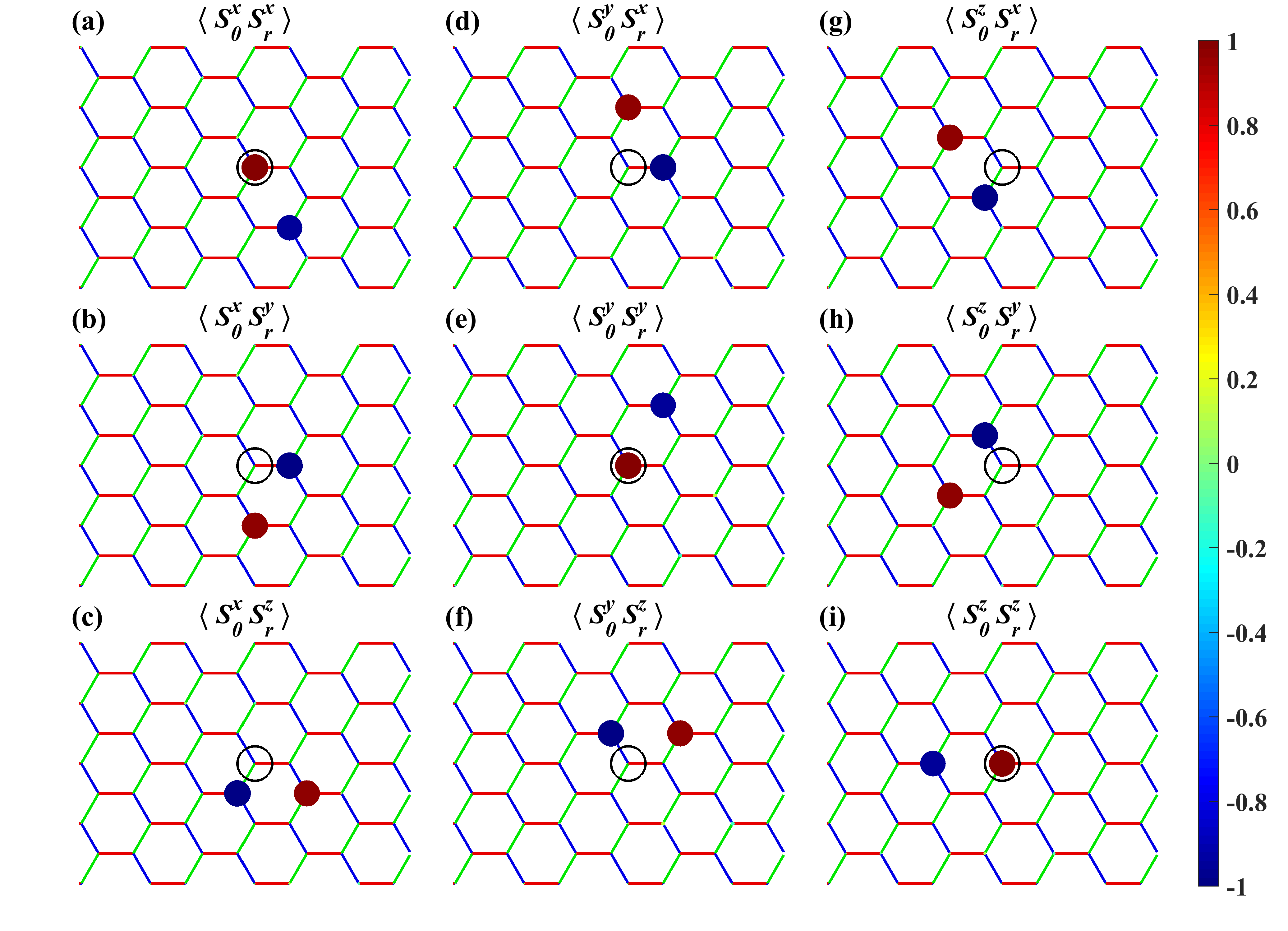}
  \caption{Static spin-spin correlation function,
    $\tilde{S}^{\alpha\beta}=\left\langle S^{\alpha }_{\bm o }S^{\beta
      }_{\bm r}\right\rangle =\ \int{S^{\alpha\beta}\left({\bm r},\omega
      \right)d\omega }$ mapped on the real-space lattice with respect to
    an arbitrary origin denoted by the cross ($\times $).  The
    temperature of the simulation is $T= 10^{-5} \Gamma$.}
  \label{fig04}
\end{figure*}

\begin{figure}[t]
  \includegraphics[clip,width=1.0\columnwidth]{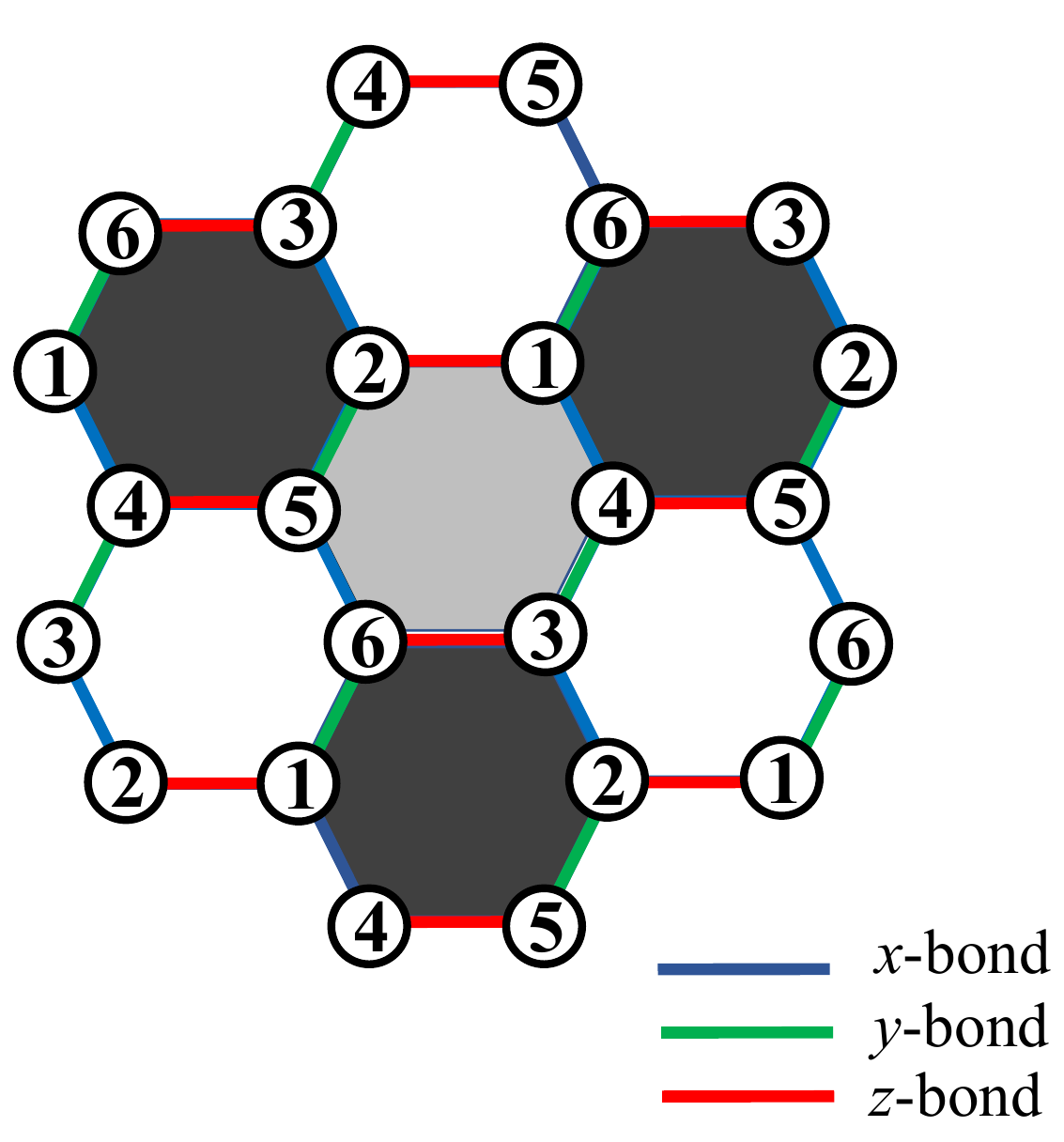}
  \caption{Six sublattice decomposition of the honeycomb lattice. We
    also show the three different decompositions of the lattice into
    isolated hexagonal plaquettes.}
  \label{symm}
\end{figure}

A similar behavior is observed in the static real space spin-spin
correlation function, $\tilde{S}^{\alpha\beta} ({\bm r})=\left\langle
  S^{\alpha }_{\bm 0 }S^{\beta }_{\bm r}\right\rangle =\
\int{S^{\alpha\beta}\left({\bm r},\omega \right)d\omega }$, shown in
Fig.~\ref{fig04}. This figure also includes off-diagonal components of
the spin-spin correlation function.  In all the cases, the correlation
function is significant only for distances $|{\bm r}|$ equal or
smaller than the separation between third nearest neighbors (opposite
sites of each hexagon).  Moreover, a subset of the nine
spin-spin correlator components $\alpha\beta$ vanishes, for any given
$\bm r$. The form of the real space correlations, as depicted in
Fig.~\ref{fig04}, is well accounted for by considering the symmetries
of the Hamiltonian, Eq. \ref{gammamodel}. To show this, we begin by
considering the three ways in which one can decompose the honeycomb
lattice into hexagon plaquettes, shown in Fig. \ref{symm}. With each
plaquette of a given decomposition we associate six spin components,
one from each site around the plaquette. Specifically, a spin
component is associated with a neighboring hexagon plaquette if it is
of the same type as the bond connecting it with a neighboring
plaquette of the same decomposition.  There exist three symmetry
operations~\cite{rousochatzakis_classical_2017}, one for each plaquette
decomposition, which correspond to $\pi$ spin rotations about an axis that depends on
the sublattice which each spin belongs to. For example, the
symmetry operation corresponding to the white plaquettes in
Fig. \ref{symm}, corresponds to a $\pi$-rotation about the $x$-axis for sublattices 1 and 4,
about the $y$-axis for sublattices 2 and 5 and about the $z$-axis for sublattices 3 and 6.
This transformation leaves the spin components $\{S_1^x, S_2^y, S_3^z, S_4^x, S_5^y,
S_6^z\}$ invariant (the subscript is the  six
sublattice index), while it changes the sign of the other ones. In appendix \ref{sec:symm}, we show that these
symmetries, which are also symmetries of the quantum model, result in
vanishing correlations between spin components which are associated
with plaquette of different decompositions, according to the rule we
just described.

 However, the spin-spin correlations are further restricted in the
classical limit. Classically, the three components of each spin
commute with each other, and as a result, one can define a local transformation that flips the sign of
an individual spin component. The classical version of the $\Gamma$ model is invariant under 
a {\it local} symmetry transformation that changes the sign of
one spin component of each of the six spins in a  {\it single} hexagon plaquette.
The spin component that changes  sign is the one corresponding to the only bond which does not connect 
two spins in the same hexagon~\cite{rousochatzakis_classical_2017}. For instance, the symmetry transformation changes the sign of $\{S_1^x, S_2^y, S_3^z, S_4^x, S_5^y, S_6^z\}$ for a {\it single} white hexagon plaquette. 
This is the local symmetry that gives rise to the macroscopic degeneracy of the classical ground state
manifold. Consequently, the correlation function $\left\langle
S^{\alpha }_{\bm 0 }S^{\beta }_{\bm r}\right\rangle$ vanishes unless
both $S_0^\alpha$ and $S_{\bm r}^\beta$ belong to the same
single hexagon. This restricts correlations to third neighbor at most,
and determines the specific components which have non zero
correlations, as seen in Fig. \ref{fig04}. For example, only the
diagonal components of the on-site correlations are non-zero, since
different components of a given spin are associated with different
plaquettes, and thus, uncorrelated.  Similarly, the only other
non-vanishing diagonal components of the spin-spin correlation
function appears for third neighbors, e.g., for the white plaquette
mentioned earlier we have only a finite $\left\langle S_1^xS_4^x
\right\rangle$ in addition to the on-site correlation. Similar
considerations restrict finite off-diagonal correlations to nearest and
second nearest neighbors around one plaquette.

\begin{figure}[htp]
  \centering
  \includegraphics[trim={0cm 0cm 0cm 0cm},clip,width=1.0\columnwidth]{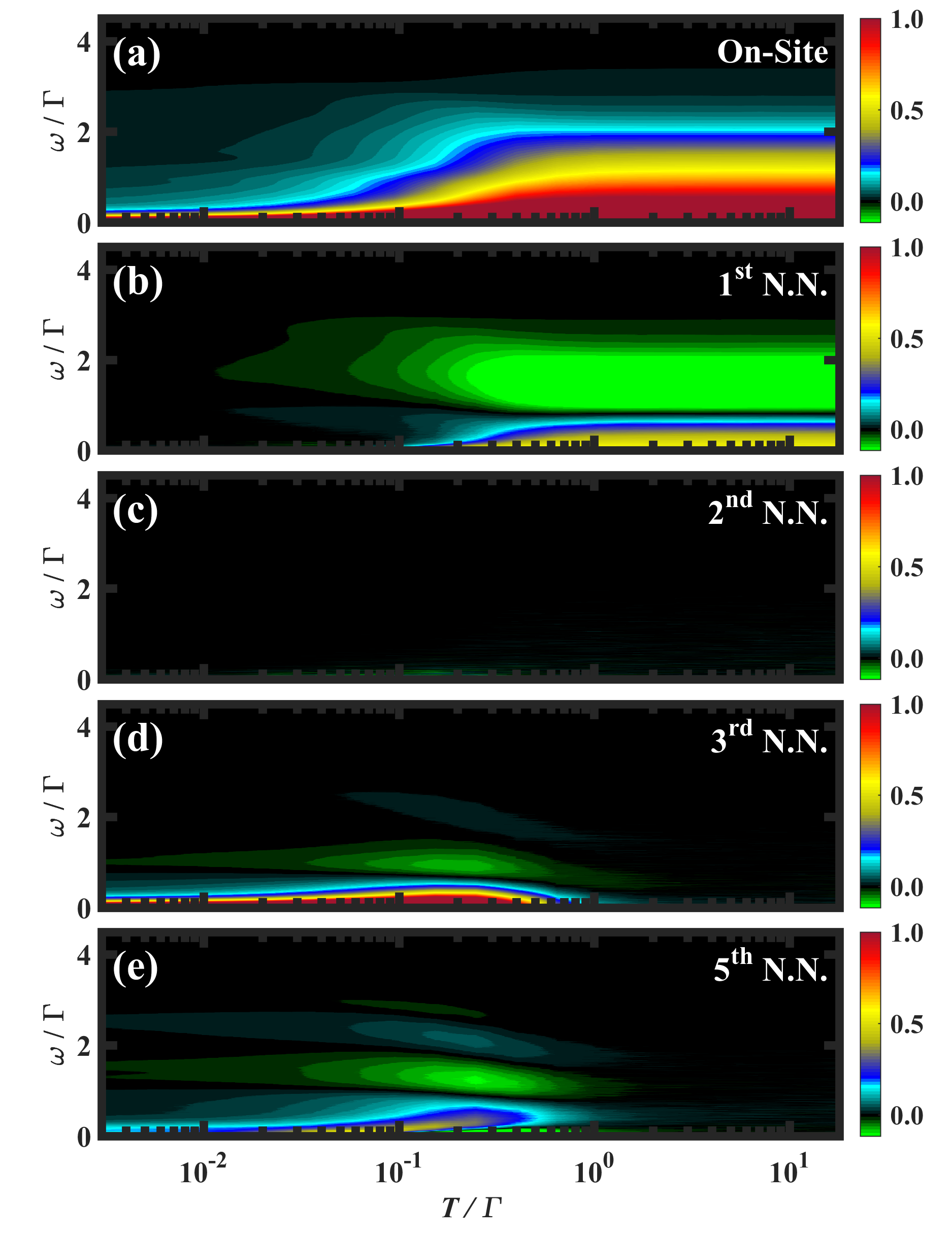}
  \caption{$S\left({\bm r},\omega \right)$ as a function of
    temperature and energy ($\hbar \omega $)for AFM ($\Gamma>0$) Gamma
    model.  Panels (a), (d) and (e) show $-S\left({\bm r},\omega
    \right)$ for ${\bm r}$ connecting 1st, 3rd and 5th
    nearest-neighbor sites, respectively. Panel (b) shows $S\left({\bm
        r},\omega \right)$ for ${\bm r}$ connecting nearest-neighbor
    sites. Panel (c) shows that $S\left({\bm r},\omega \right)=0$ for
    ${\bm r}$ connecting second nearest neighbor sites.}
  \label{fig05}
\end{figure}

\begin{figure}[htp]
  \centering
  \includegraphics[clip,width=\columnwidth]{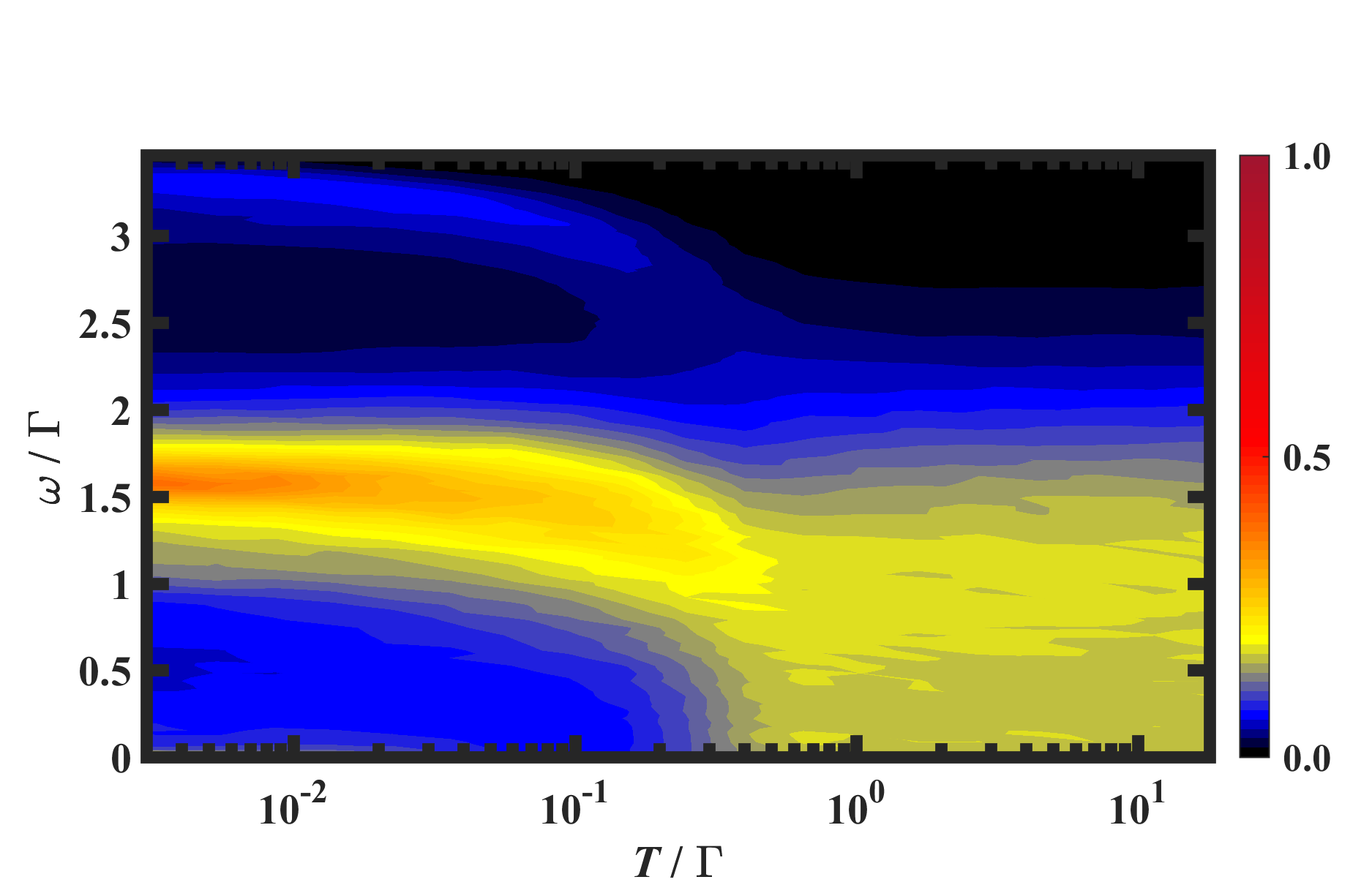}
  \caption{Temperature evolution of $S ({\bm Q} ={\bm \Gamma
    },\omega)$ for AFM ($\Gamma>0$) Gamma model. }
  \label{fig06}
\end{figure}

 Fig.~\ref{fig05} shows the temperature and frequency
dependence of $S\left({\bm r} ,\omega \right)$ for several values of
${\bm r}$.  In agreement with the symmetry analysis given in Appendix~\ref{sec:symm},
$S\left({\bm r} ,\omega \right)$ vanishes for any frequency when ${\bm r}$ connects second
nearest neighbor sites [see Figure~\ref{fig05}~(b)]. The temperature dependence of $S\left({\bm r} ,\omega \right)$
for other values of ${\bm r}$ indicates a crossover from partially
precessional to fully diffusive behavior at a temperature of order
$\Gamma$.  The temperature dependence of $S\left( {\bm Q} ={\bm \Gamma
  },\omega \right)$ shown in Figure~\ref{fig06} confirms this
crossover, indicating that the system evolves continuously from a
low-temperature ($T< \Gamma$) correlated liquid (classical $\Gamma$
liquid) to a high-temperature paramagnetic state.

\section{Langevin dynamics}
\label{sec:lan}

Although, as noted in the previous section, neutron scattering
experiments are faster than spin-lattice relaxation, a Langevin
approach may still be used to analytically understand dynamic
correlations in LL dynamics. The non-linear nature of
Eq. (\ref{eq:LL}) gives rise to strong relaxation due to inelastic
processes, in addition to the more direct precessional dynamics. In
other words, the fluctuating spins act as their own heat bath --
leading to strong relaxation. At a phenomenological level, it is
possible to capture the two types of spin dynamics in a generalized
Langevin equation, where a linear term describes the spin relaxation,
while precession is included in a residual non-linear term.  We thus
propose to study the stochastic dynamics of a system of soft classical
spins~\cite{Colon09} $S_i^\alpha$, given by
\begin{equation}
  \label{eq:lan}
  \frac{\partial S_i^\alpha}{\partial t} = g\epsilon_{\alpha\beta\gamma}S_i^\beta
  \frac{\partial H}{\partial S_i^\gamma} - \gamma\frac{\partial{H}}
  {\partial S_i^\alpha} + \eta_i^\alpha,
\end{equation}
where $i,j$ denote a site on a honeycomb lattice, $\alpha,\beta=x,y,z$ and
the noise term $\eta_i^\alpha$ obeys
\begin{equation}
  \label{eq:noise}
  \left\langle\eta_i^\alpha(t)\eta_j^\beta(t')\right\rangle 
  = 2\gamma T \delta_{ij}\delta_{\alpha\beta}\delta(t-t').
\end{equation}
As described below, the unitless phenomenological parameters --
$\gamma$ setting the relaxation rate and $g$ the precession rate --
are chosen by comparing the dynamical correlations with the results of
the LL simulation, while $T$ is simply the temperature.  The spins are
taken to be soft with mass $\Delta$, and with a general nearest
neighbor spin interaction, $K_{ij}^{\alpha\beta}$, as given by the
following Hamiltonian
\begin{equation}
  \label{eq:H}
  H = \sum_{\alpha\beta}\sum_{\braket{ij}}K_{ij}^{\alpha\beta}
  S_i^\alpha S_j^\beta + \frac{\Delta}{2}\sum_{i\alpha}\left(S_i^\alpha\right)^2,
\end{equation}
The mass $\Delta$ is an additional, temperature dependent,
phenomenological parameter which determines the average spin size
$\braket{S_i^2}$.  Quadratic Hamiltonians of this type are suitable
for studying spin dynamics in cases where there is no long range
order. We will apply this to the classical $\Gamma$ model, which has
macroscopic degeneracy in its ground state, preventing long range
order even at low temperatures. A similar treatment for the Kitaev
model is given in the Appendix. Following a path integral formulation
of the MSR approach\cite{martin_statistical_1973,
  de_dominicis_field-theory_1978, wachtel_transverse_2014}, we write a
generating functional for dynamical correlations as
\begin{equation}
  \label{eq:Z1}
  Z\!=\!\left\langle\!\int DS\,|M|\,
    \delta\!\left(\frac{\partial S_i^\alpha}{\partial t}\!-\! 
      g\epsilon_{\alpha\beta\gamma}S_i^\beta \frac{\partial H}{\partial S_i^\gamma} 
      \!+\!\gamma\frac{\partial{H}} {\partial S_j^\alpha} 
     \!-\! \eta_i^\alpha\right)\!\right\rangle_\eta \!,
\end{equation}
where $\braket{\cdots}_\eta$ denotes averaging over the noise
fluctuations $\eta$, and $M$ is a Jacobian matrix. For the model we
consider here we may take the determinant $|M|=1$. Writing the delta
function as an integral over $\hat S_i^\alpha$, and averaging over
$\eta$, we obtain
$Z = \int D\hat S\,DS\,e^{-\mathcal{S}}$ where the MSR action is given
by
\begin{eqnarray}
  \label{eq:S1}
  \mathcal{S} & = & -\int dt\left[\hat S_i^\alpha\left(\partial_tS_i^\alpha
      +\gamma K_{ij}^{\alpha\beta}    S_j^\beta+\gamma\Delta S_i^\alpha\right)
    \right. \nonumber \\ & & \left.\qquad\qquad +\gamma T (\hat S_i^\alpha)^2
    -g\epsilon_{\alpha\beta\gamma}\hat S_i^\alpha S_i^\beta K_{ij}^{\gamma\delta} 
    S_j^\delta\right],
\end{eqnarray}
Within this formalism it is possible to calculate dynamical
correlation functions, using perturbation theory in $g$.

\begin{figure}[bb]
  \centering
  \includegraphics[width=\linewidth]{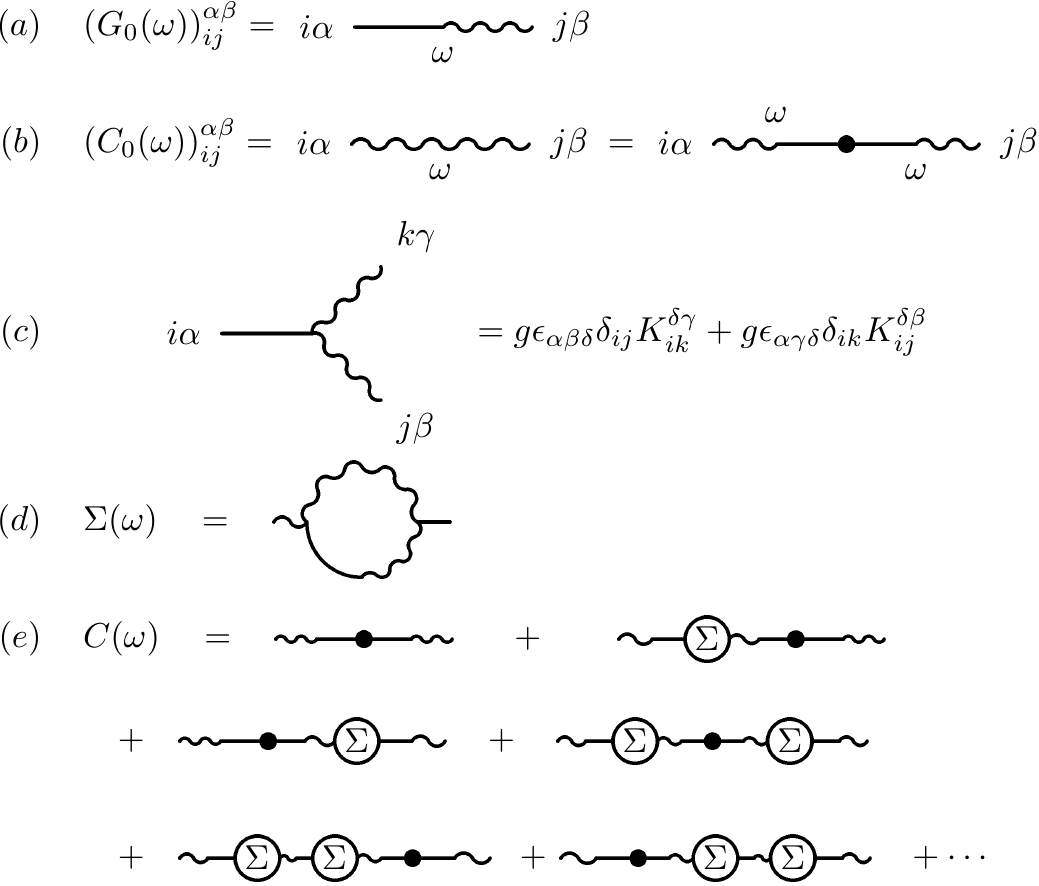}
  \caption{Feynman diagrams for the MSR formalism: (a) Bare response
    function. (b) Bare correlation function. (c) Precession
    vertex.  (d) Self-energy diagram. (e) Perturbative
      expansion of the dressed correlation function, in powers of the
      self-energy $\Sigma$, see Eq. (\ref{eq:C}).}
  \label{fig:diagrams}
\end{figure}
\emph{Zeroth order in} $g$: When $g=0$, the dynamics is purely
relaxational. The bare response Green's function, defined as
\begin{equation}
  \label{eq:G0}
  \left(G_0(\omega)\right)_{ij}^{\alpha\beta}\equiv\left\langle \hat 
    S_i^\alpha(-\omega)S_j^\beta(\omega)    \right\rangle_0,
\end{equation}
is given by (its inverse)
\begin{equation}
  \label{eq:G0inv}
  \left(G^{-1}_0(\omega)\right)_{ij}^{\alpha\beta} = (-i\omega+\gamma\Delta)
  \delta_{\alpha\beta}\delta_{ij}  + K_{ij}^{\alpha\beta}
\end{equation}
We will represent this diagrammatically using Fig. \ref{fig:diagrams}a.
Similarly, the bare correlation function is given by
\begin{equation}
  \label{eq:C0}
  \left(C_0(\omega)\right)_{ij}^{\alpha\beta} \equiv \left\langle 
    S_i^\alpha(-\omega)S_j^\beta(\omega)\right\rangle_0 = 2\gamma T \left(
    G_0^\dagger(\omega)G_0(\omega)\right)_{ij}^{\alpha\beta}
\end{equation}
Diagrammatically, this is represented in Fig. \ref{fig:diagrams}b
where the noise vertex is represented by a dot, $\bullet = 2\gamma T$.

In the pure $\Gamma$ model, which is defined by
\begin{equation}
  \label{eq:Kitaev}
  K_{ij}^{\alpha\beta}=\left\{
    \begin{array}{cc}
      \Gamma & \alpha\ne\beta\ne\braket{ij} \\ 0 & {\rm otherwise}
    \end{array}\right. ,
\end{equation}
the classical degrees of freedom can be divided into sectors -- one
for each hexagon. For example, as discussed in the previous section,
going around one of the white hexagons in fig. \ref{symm}, we can
identify six spin components which interact only within
themselves. Their dynamics is independent of the rest of the system,
manifesting the macroscopic degeneracy of the classical system. We
rename these degrees of freedom as follows,
\begin{equation}
  \label{eq:sigmas}
  \left\{ S_1^x, S_2^y, S_3^z, S_4^x, S_5^y, S_6^z\right\} \equiv \left\{\sigma_1, \sigma_2,
    \sigma_3, \sigma_4, \sigma_5, \sigma_6\right\}.
\end{equation}
The Hamiltonian, restricted to this sector, is given by
\begin{equation}
  \label{eq:Hsigma}
  H_{\rm hexagaon} = \sum_{l=1}^6 \left(\Gamma \sigma_l\sigma_{l+1} + \frac{\Delta}{2}
    \sigma_l^2\right).
\end{equation}
The energy eigenvalues, which determine the relaxation rate, are
$\varepsilon_m = 2\Gamma\cos{\pi m}/{3}+\Delta,\quad m=1\dots 6$.
Clearly, the model is stable only for $\Delta>2|\Gamma|$.  The $g=0$
dynamical correlation function within these six spin components is
given by
\begin{equation}
  \label{eq:hexcor}
  \left\langle \sigma_l(\omega)\sigma_{l'}(\omega')\right\rangle
  \!=\! \frac{1}{6}\!\sum_{m=1}^6\!\cos\!\left(\!\frac{\pi m(l-l')}{3}\!\right)
  \frac{2\gamma T\,2\pi\delta(\omega+\omega')}
  {\omega^2 + \gamma^2\varepsilon_m^2},
\end{equation}
while the correlation with spin components which do not appear in
Eq. (\ref{eq:sigmas}) is identically zero.  $\Delta$ may be chosen so
that $\braket{\sigma_l^2}$ is a constant; at the $g=0$ level $\Delta$
should obey $\braket{\sigma_l^2}=\sum_m(T/6\varepsilon_m)=1$. Clearly,
the dynamics for $g=0$ is purely relaxational, with a peak at
$\omega=0$. Thus, $\gamma$ can be chosen such that
  Eq. (\ref{eq:hexcor}) reproduces the width of the low energy peak as
  obtained in the LL simulations. Focusing on the dynamic structure
factor $\braket{S^\alpha({\bf q},\omega)S^\alpha(-{\bf q},-\omega)}$,
we note that for each hexagon $\braket{S^x_lS^x_{l'}}\ne 0 $ only for
$l,l'=1,4$, $\braket{S^y_lS^y_{l'}}\ne 0$ only for $l,l'=2,5$ and
$\braket{S^z_lS^z_{l'}}\ne 0$ only for $l,l'=3,6$, in line with the
symmetry consideration of the previous section and in Appendix
\ref{sec:symm}. In the anti-ferromagnetic $\Gamma$ model, the lowest
eigenvalues is $\varepsilon_3=-2\Gamma+\Delta$. Noting that there are
non-zero correlations only for $l-l'=0,3$, we find that the
contribution of this eigenvalue to the spectrum at $\bm{\Gamma}$
vanishes, and one would expect only the higher energy modes,
i.e. faster relaxations, to contribute at this momentum. A similar
argument holds for the low energy correlations at $\bm{X}$. Thus we
find a depletion in the dynamic structure factor at $\bm{\Gamma}$ and
$\bm{X}$, echoing the analysis of the zero modes in the previous
section.

\begin{figure}[tt]
  \centering
  \includegraphics[width=\linewidth]{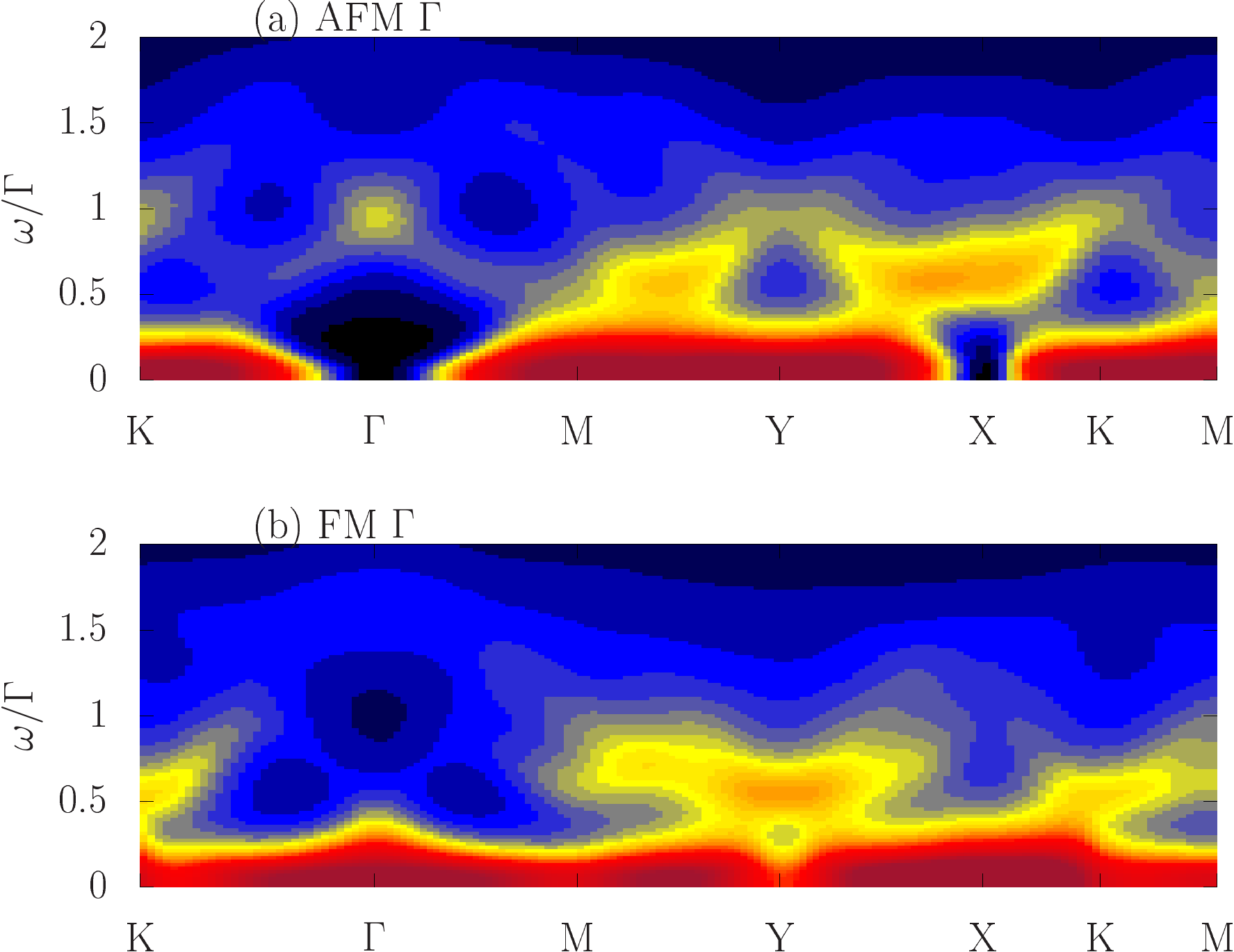} 
  \caption{Dynamic structure factor as obtained from Eq. (\ref{eq:C}),
    for (a) the AFM $\Gamma>0$, and (b) the FM $\Gamma<0$ models. Here
    we have used $\gamma=0.12$, $\Delta=2.05\Gamma$ and
    $g^2T=0.04\Gamma$.}
  \label{fig:msr_gamma}
\end{figure}
\emph{Perturbation theory in $g$.} It is possible to treat the
precession term using diagrammatic perturbation theory. We represent
the \emph{symmetrized} precession vertex by Fig. \ref{fig:diagrams}c.
The dressed correlation function, $C_{ij}^{\alpha\beta}(\omega)$, can
be calculated approximately, by summing over a subset of infinite
diagrams, as shown in Fig. \ref{fig:diagrams}. Specifically, we
approximate $C(\omega)$ by a product of two infinite series,
\begin{eqnarray}
  \label{eq:C}
  C(\omega)\approx2\gamma T\left|G_0(\omega)\sum_{n=0}^\infty 
    \left(\Sigma(\omega)G_0(\omega)\right)^n\right|^2,
\end{eqnarray}
where the `self energy' $\Sigma$ is calculated to leading order in
$g$.  The self energy term for the $\Gamma$ model dynamics mixes
different sectors, and therefore its calculation must be done in
Fourier space. However, the procedure is no different than in quantum
field theory, once the appropriate Feynman rules are
determined. Fig. \ref{fig:msr_gamma} shows the resulting dynamic
structure factor. Eq. (\ref{eq:hexcor}), obtained for $g=0$,
qualitatively accounts for the low frequency features, including the
depletion at $\bm{\Gamma}$ and $\bm{X}$ in the AFM case. The main
qualitative effect of finite $g>0$ on the dynamic structure factor, is
the addition of correlations peaked at finite frequency, due to the
precession of the spins. In Appendix \ref{sec:msrkitaev} we describe
the calculation for the Kitaev model, which is simpler, and can be
done in real space.  Furthermore, the closed form result for the
Kitaev model shows that the self energy is larger for the mode which
is suppressed at low energies. Similar behavior is observed in
Fig. \ref{fig:msr_gamma} for the $\Gamma$ model, where the precession
features appear at finite frequency at the same momentum positions
with depleted low energy correlations. Besides the qualitative effect
of precessional dynamics, a finite $g$ is also expected to renormalize
the values of $\gamma$ and $\Delta$ required to fit the numerical data
obtained at a given temperature.

\section{Dynamics of the spin-$\frac{1}{2}$ model}
\label{sec:ed}

\begin{figure}[tth]
\centering
\includegraphics[width=4.0cm]{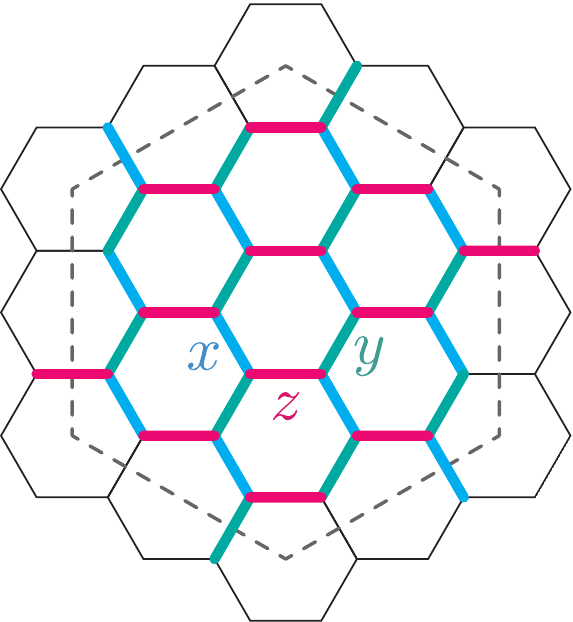}
\caption{(color online):
Finite-size honeycomb clusters with 24 spins and periodic boundary conditions.
Bonds along the three different directions are labeled as
the $x$, $y$, and $z$ bond, which are along -60$^{\circ}$, 60$^{\circ}$,
and horizontal directions, respectively.
}
\label{fig_honeycomb_18_24}
\end{figure}

\begin{figure}[hbt]
\centering
\includegraphics[width=0.5\textwidth]{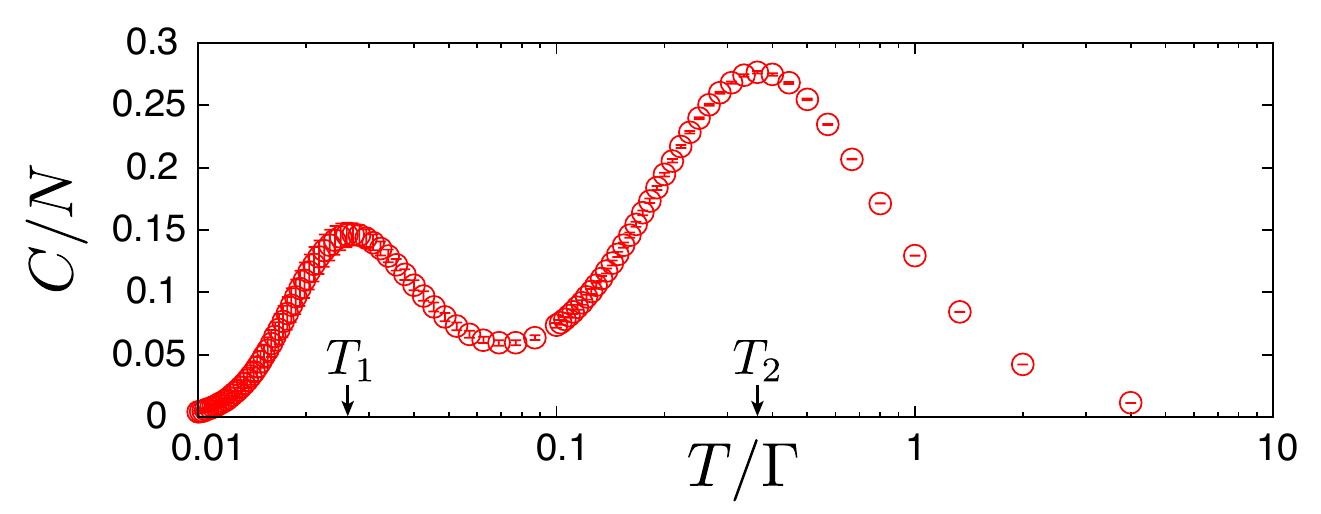}
\caption{(color online):
Specific heat $C/N$
of the AFM $\Gamma$ model on the $N=24$ site cluster~\cite{catuneanu_path_2018},
calculated by the typical pure quantum states approach~\cite{PhysRevE.62.4365,PhysRevLett.111.010401}.
There are two maxima in the temperature dependence of $C/N$. 
The error bars
are the standard errors estimated by 32 random initial vectors.
}
\label{FigC_S1half}
\end{figure}

\begin{figure*}[hbt]
\centering
\includegraphics[width=1.05\textwidth]{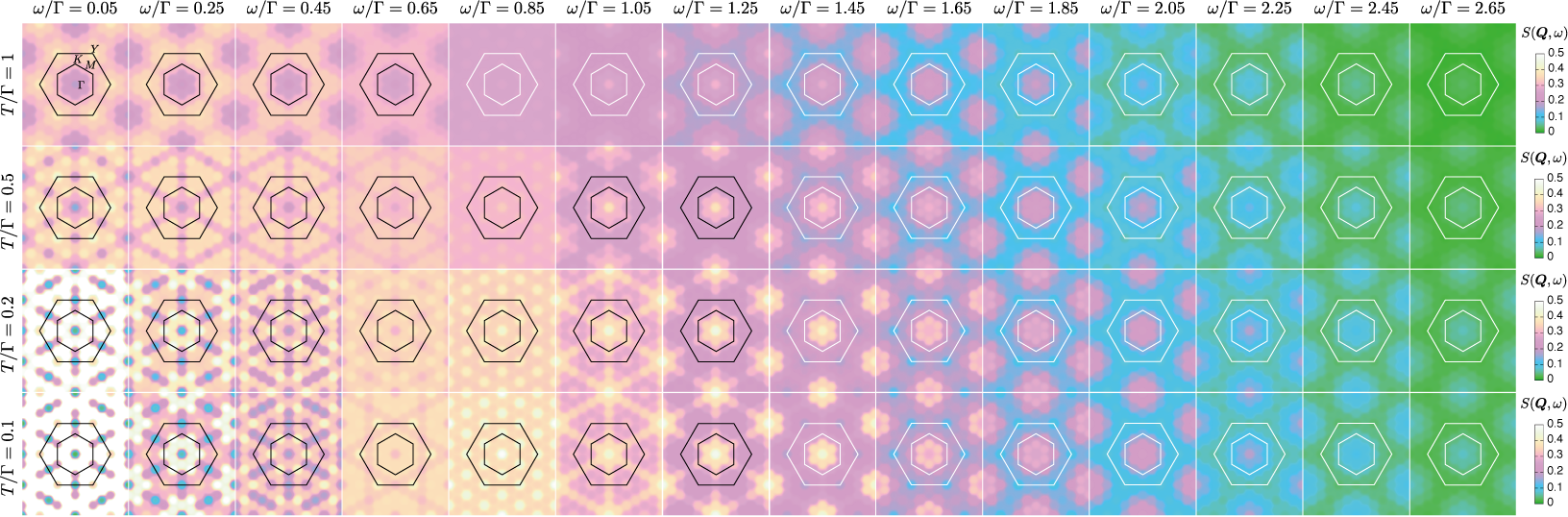}
\caption{(color online):
Equi-energy slices of the dynamical spin structure factors of the $\Gamma>0$ model.
The momentum dependence of the equi-energy slices is shown by changing temperature and frequency.
From the top row to the bottom row, the equi-energy slices of the dynamical spin structure factors
are shown at $T/\Gamma=1$, $0.5$, $0.2$, and $0.1$.  
The equi-energy slices are prepared by averaging the spectra within an energy window whose width is 0.1.
For visibility, the dynamical spin structure factors at discrete momenta obtained by the simulation
are interpolated. Here, the broadening factor $\eta/\Gamma =0.02$ is used (see Appendix \ref{details_of_ED} for the definition of $\eta$).
}
\label{FigSQomegaBZ}
\end{figure*}

\begin{figure}[hbt]
\centering
\includegraphics[width=0.4\textwidth]{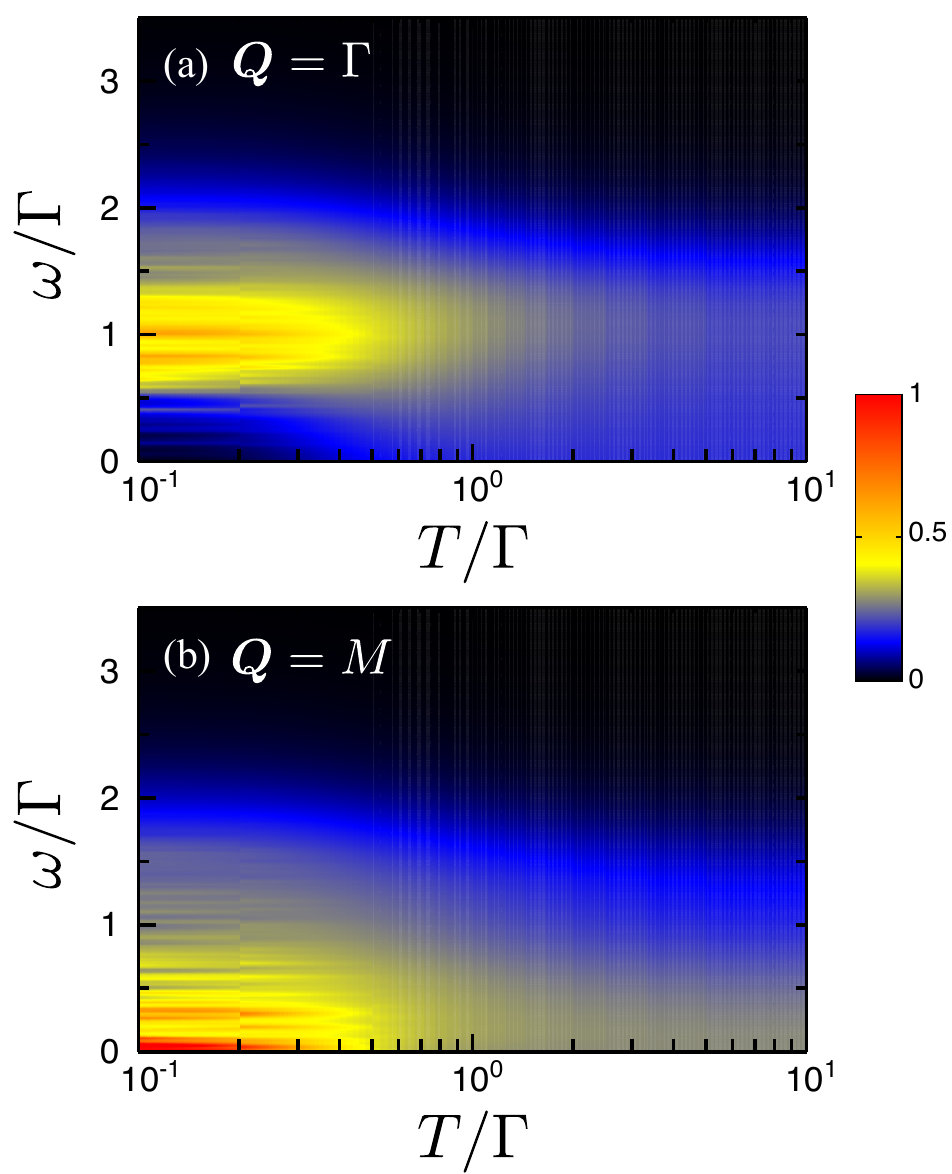}
\caption{(color online): Temperature evolution of
  $S(\mbox{\boldmath$Q$}=\Gamma,\omega)$ and
  $S(\mbox{\boldmath$Q$}=M,\omega)$ for the $\Gamma>0$ model.  }
\label{FigGammaM}
\end{figure}

\begin{figure}[hbt]
\centering
\includegraphics[width=0.4\textwidth]{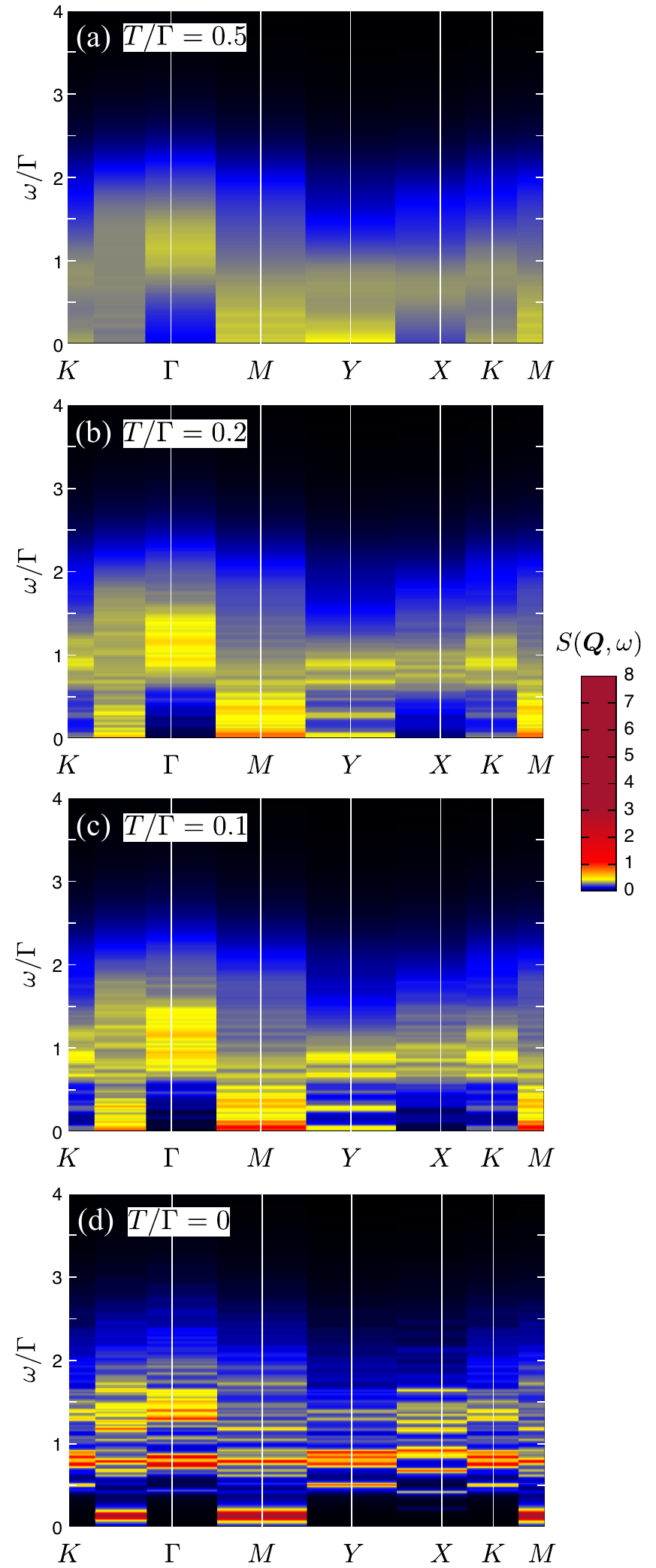}
\caption{(color online):
Dynamical spin structure factors of the $S=1/2$ AFM $\Gamma$ model at
(a) $T=0.5$, (b) $T=0.2$, (c) $T=0.1$, and (d) $T=0$,
along symmetry lines.
For visibility, the broadening factor $\eta/\Gamma =0.02$ is used (see Appendix \ref{details_of_ED} for the definition of $\eta$).
}
\label{Figdiagram}
\end{figure}

\begin{figure*}[hbt]
\centering
\includegraphics[width=0.6\textwidth]{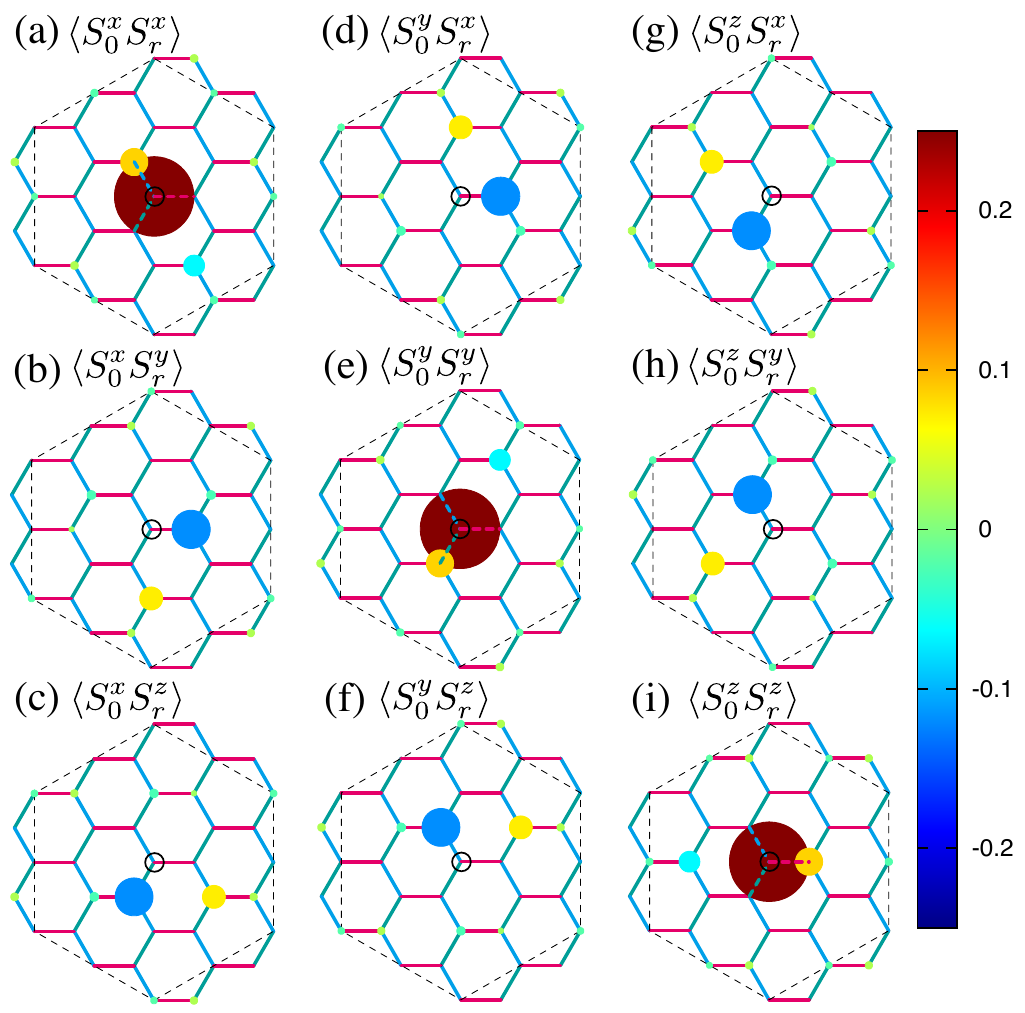}
\caption{(color online):
Real-space static spin-spin correlation function, $\langle {S}^{\alpha}_{\mbox{\boldmath$0$}} {S}^{\beta}_{\mbox{\boldmath$r$}}\rangle$ $(\alpha,\beta = x,y,z)$,
of the $S=1/2$ $\Gamma>0$ model on the 24 site cluster
with periodic boundary condition, at $T=0$.
The location of the origin $\mbox{\boldmath$0$}$ is denoted by the open circle ($\circ$).
The radiuses of the closed circles at $\mbox{\boldmath$r$}$ represent the amplitude of $|\langle {S}^{\alpha}_{\mbox{\boldmath$0$}} {S}^{\beta}_{\mbox{\boldmath$r$}}\rangle|$,
while the color of the closed circles shows $\langle {S}^{\alpha}_{\mbox{\boldmath$0$}} {S}^{\beta}_{\mbox{\boldmath$r$}}\rangle$.
Within numerical accuracy, there is no spin-spin correlation represented by a circle with a radius smaller than the width of the solid lines representing the bonds.
}
\label{FigS0SR_S1half}
\end{figure*}

\begin{figure}[hbt]
\centering
\includegraphics[width=0.5\textwidth]{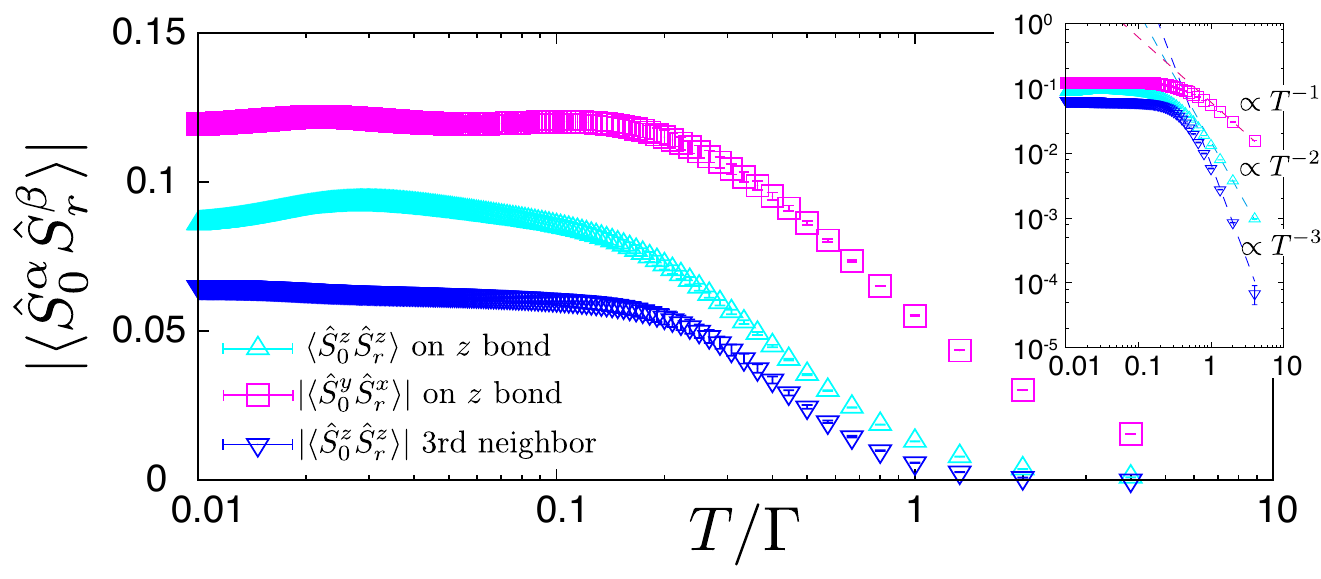}
\caption{(color online):
Temperature dependence of static spin-spin correlations of the AFM $\Gamma$ model on the 24 site cluster.
Here, the error bars, which are smaller than or comparable to the symbol size,
are the standard errors estimated by several random initial vectors.
}
\label{FigfTSS_S1half}
\end{figure}

Using the exact-diagonalization method, described in Ref.~\onlinecite{yamaji2018numerical} and Appendix \ref{details_of_ED}, we study
the finite-temperature dynamical spin structure factors
of the $S=1/2$ AFM $\Gamma$ model.
Here, we use a 24 site cluster with periodic boundary conditions,
illustrated in Fig. \ref{fig_honeycomb_18_24}.  As explained below,
the dynamical spin structure factor of the quantum model shows a
gradual classical-to-quantum crossover when the temperature is
decreased.

Before going into details of the
quantum spin dynamics of the $\Gamma$ model,
we summarize the energy scale of the quantum $S=1/2$
$\Gamma$ model.
As shown in Ref.~\onlinecite{catuneanu_path_2018} and Fig.~\ref{FigC_S1half}, there
are two temperature scales given by the peaks in the temperature
dependence of the specific heat.  The higher-temperature peak appears
around $T_2/\Gamma \sim 0.4$ and the lower-temperature peak emerges below
$T_1/\Gamma \lesssim 0.03$.
The two peak structure in the temperature
dependence of the specific heat has been found in the Kitaev
model~\cite{PhysRevB.92.115122} and in the proximity of the Kitaev's
spin liquid~\cite{PhysRevB.93.174425}, although the balance of the
entropy released by these two peaks is different from that of the
$\Gamma$ model.

In Fig.~\ref{FigSQomegaBZ}, the momentum dependence of the equi-energy
slices of $S(\mbox{\boldmath$Q$},\omega)$ are shown by changing
temperature and frequency.
The equi-energy slices are prepared by
averaging the spectra within an energy window whose width is 0.1.
The momentum dependence is numerically interpolated for visibility without
changing the simulation results at the discrete momenta
$\mbox{\boldmath$Q$}$ compatible with the finite size cluster.
At $T/\Gamma = 1 > T_2/\Gamma \sim 0.4$, almost momentum-independent
behaviors of $S(\mbox{\boldmath$Q$},\omega)$ are found except around $\Gamma$ point,
where spectral weight is suppressed for $\omega/\Gamma < 1$ and is shifted to the high energy region $\omega/\Gamma > 1$.
Below $T/\Gamma = 0.5 \sim T_2/\Gamma$, the suppression of the low-energy spectral weight (or relaxational dynamics) at $\Gamma$ and $X$ points
becomes notable, which is consistent with the classical dynamics.

To examine the temperature dependence of the low energy spectral
weight, we show the temperature evolution of
$S(\mbox{\boldmath$Q$}=\bm\Gamma,\omega)$ in comparison with
$S(\mbox{\boldmath$Q$}=\bm M,\omega)$ in Fig.\ref{FigGammaM}.  Below
the high temperature scale $T/\Gamma\sim 0.4$,
$S(\mbox{\boldmath$Q$}=\bm \Gamma,\omega)$ shows reduced spectral
weight in the low energy region below $\omega/\Gamma \sim 0.5$, while
$S(\mbox{\boldmath$Q$}=\bm M,\omega)$ shows substantial spectral
weight in the low energy region. We further note that the Fourier
transformation of $S(\mbox{\boldmath$Q$},\omega)$,
$S(\mbox{\boldmath$r$}_{\ell m},\omega)$, also satisfies the symmetry
properties at finite temperatures, as discussed in section
\ref{sec:LL}.

For closer comparison with the classical dynamics, we show 
$S(\mbox{\boldmath$Q$},\omega)$ at $T/\Gamma=0.5$, $0.2$, $0.1$, and $0$ along symmetry lines
in Fig.\ref{Figdiagram}.
In addition to the suppression of the low-energy spectral weight at $\Gamma$ and $X$ points,
which is in common with the classical spin dynamics,
the low-energy spectral weight for $\omega/\Gamma \lesssim 0.5$ decreases at $K$ and $Y$ upon cooling.
This suppression at the $K$ and $Y$ points is characteristic of the $S=1/2$ $\Gamma$ model.

To gain insight into the difference between the
  quantum and classical dynamics at the $K$ and $Y$ points, we examine
  the static spin-spin correlation function at zero temperature.  In
  Fig.\ref{FigS0SR_S1half}, $\langle
  {S}^{\alpha}_{\mbox{\boldmath$0$}}
  {S}^{\beta}_{\mbox{\boldmath$r$}}\rangle$ $(\alpha,\beta =
  x,y,z)$ are shown in the 24 site cluster with the periodic boundary
  condition.  These correlators in the quantum model are very similar
  to the static correlation functions of the classical model, as shown
  in Fig.~\ref{fig04}.  Due to the symmetry of the $\Gamma$ model, the
  static spin-spin correlation functions are zero for many spin pairs.
  However, there are differences among them: For example, there exist
  finite nearest-neighbor correlations $\langle
  {S}^{\alpha}_{\mbox{\boldmath$0$}}
  {S}^{\beta}_{\mbox{\boldmath$r$}}\rangle $ $(\alpha,\beta =
  x,y,z)$ for $\alpha = \beta$ and the additional second
  nearest-neighbor correlations for $\alpha \neq \beta$. The fact that
  these correlations are finite, while they are absent in the
  classical limit, indicates that quantum fluctuations are important
  even when $T>T_1$.


The most significant difference is the nearest-neighbor correlations for $\alpha=\beta$, which are  zero
in the classical model  due to the  the local symmetry discussed in Sec.~\ref{sec:LL}.
The nearest-neighbor ferromagnetic correlations in the real space
hinder the antiferromagnetic low-energy fluctuations at the $Y$ and $K$ points in the momentum space.
As a result, these correlations harden the spin fluctuations at these momenta.
In other words, these correlations suppress the relaxational dynamics and introduce the quasi-collective precessional dynamics
at these momenta.

A quantitative description of the classical-quantum crossover is
obtained by examining temperature dependence of the typical static
spin-spin correlation functions shown in Fig.~\ref{FigfTSS_S1half}.
While the off-diagonal nearest-neighbor correlations $\langle
{S}^{\alpha}_{\mbox{\boldmath$0$}}
{S}^{\beta}_{\mbox{\boldmath$r$}}\rangle$ on the $\gamma$ bond,
where $(\alpha,\beta,\gamma)$ is a permutation of $(x,y,z)$, are
dominant at temperatures above and around $T_2/\Gamma \sim 0.4$, the
diagonal nearest-neighbor correlations $\langle
{S}^{\alpha}_{\mbox{\boldmath$0$}}
{S}^{\alpha}_{\mbox{\boldmath$r$}}\rangle$ start to saturate upon
cooling for $T<T_2$.  Therefore, the classical precessional dynamics
due to the off-diagonal nearest-neighbor correlations governs the
dynamics for $T\gtrsim T_2$.  On the other hand, the emergent quantum
dynamics is generated by the diagonal nearest-neighbor correlations
for $T<T_2$.  Here, we note that dominance of the off-diagonal
nearest-neighbor correlations originates from their Curie-like
temperature dependence in the high temperature region, which is
evident in the inset of Fig.~\ref{FigfTSS_S1half}, while temperature
dependence of the diagonal nearest-neighbor and third nearest-neighbor
correlations shows $T^{-2}$ and $T^{-3}$ scaling, respectively. In
Appendix \ref{sec:highT} we show how these power-law behaviors can be
obtained using a high temperature expansion.


\section{Discussion}
\label{sec:dis}

In this work, we investigated classical and quantum dynamics of the
$\Gamma$ model, the bond-dependent symmetric and anisotropic spin
interaction on the honeycomb lattice.  Such exchange interaction
arises in strongly spin-orbit-coupled Mott insulators, in addition to
the usual Heisenberg and Kitaev (the bond-dependent Ising)
interactions\cite{rau_generic_2014, rau_spin-orbit_2016}.  There exist
a number of so-called ``Kitaev materials'' such as $\alpha, \beta,
\gamma$-Li$_2$IrO$_3$ and $\alpha$-RuCl$_3$, where the Kitaev
interaction, if dominant, may lead to a quantum spin liquid phase.
However, the strength of the $\Gamma$ interaction can be as large as
that of the Kitaev interaction, especially in the case of
$\alpha$-RuCl$_3$, according to recent ab initio
computations\cite{winter_challenges_2016}.  The presence of other
interactions has been regarded as an obstacle for realizing the
quantum spin liquid ground state in this class of materials.
 
On the other hand, the $\Gamma$ interaction is also highly frustrated
at the classical level, just like the Kitaev model. Given that the
strength of this interaction is significant in some materials, the
nature of the quantum ground state of the $\Gamma$ model is highly
relevant for the interpretation of the experiments. In the case of
$\alpha$-RuCl$_3$, for example, it has been speculated that the
scattering continuum seen in recent neutron scattering experiment may
come from a nearby quantum spin liquid even though the actual ground
state is a magnetically ordered state~\cite{Banerjee16,
  Banerjee17}. The magnetic order can be suppressed by external
in-plane magnetic field and the resulting paramagnetic state is
speculated to be a field-induced quantum spin
liquid\cite{sears_phase_2017,hentrich_unusual_2018,
  wolter_field-induced_2017}.  Currently it is highly debated whether
the Kitaev interaction or other interactions or both could be
responsible for the formation of a putative quantum spin liquid ground
state.

In the present work, we focused on the $\Gamma$ interaction and
pointed out the similarity to the Kitaev model, in the correspondence
between classical and quantum dynamics. We showed that the zero mode
structure of the highly degenerate manifold of the classical ground
states is reflected in the classical dynamical spin structure factor.
In addition, we clarified different roles of relaxational and
precessional dynamics in the dynamical spin structure factor of the
classical model.  Remarkably this feature survives in the quantum
model down to very low energy scales. This would imply that the full
degenerate manifold of the classical states are participating in
quantum fluctuations down to very low energy scales.  This situation
resembles the results of the Kitaev model, obtained in a previous
study~\cite{Samarakoon17}, where the quantum dynamical spin structure
factor is qualitatively similar to the classical results down to low
energies above the small flux gap in the underlying spin liquid ground
state. This correspondence in the Kitaev model was apparent even in
the temperature/energy window where the underlying low energy degrees
of freedom are Majorana fermions, not the semiclassical spins.  Since
the quantum $\Gamma$ model is not exactly solvable, we do not know the
true quantum ground state at zero temperature.  It has been suggested
that order by quantum disorder leads to a symmetry-broken
state\cite{rousochatzakis_classical_2017}. The resemblance to the
Kitaev model, however, suggests that the quantum ground state of the
$\Gamma$ model may also be a quantum spin liquid, which results from
the ``collapse'' of the degenerate classical manifold.  Such
conclusion may also be consistent with a recent DMRG computation of
the same model\cite{gohlke_quantum_2018}, where the ground seems to be
a highly correlated quantum paramagnet.  If the $\Gamma$ model can
indeed support a quantum spin liquid ground state, the presence of
this interaction in real materials may not necessarily be an
obstruction for the realization of the quantum spin liquid ground
state. The firm answer to this question would require further studies
of the quantum and classical models with both the Kitaev and $\Gamma$
interactions.

\acknowledgments 

Work at ORNL is supported by the U.S. Department of Energy (DOE),
Office of Science, Basic Energy Sciences, Scientific User Facilities
Division. CDB acknowledges support from the Los Alamos National
Laboratory LDRD program. GW and YBK are supported by the NSERC of
Canada and the Center for Quantum Materials at the University of
Toronto. GW was additionally supported by the Israel Science
Foundation (Grant No. 585/13). YY was supported by JSPS KAKENHI (Grant
No. 16H06345) and was supported by PRESTO, JST (JPMJPR15NF).  This
research was supported by MEXT as ``Priority Issue on Post-K computer"
(Creation of New Functional Devices and High-Performance Materials to
Support Next-Generation Industries) and ``Exploratory Challenge on
Post-K computer" (Frontiers of Basic Science: Challengin the Limits).
Our numerical calculation was partly carried out at the Supercomputer
Center, Institute for Solid State Physics, University of Tokyo.  The
exact diagonalization (ED) calculations are partly double-checked by
using an open-source ED program package
$\mathcal{H}\Phi$~\cite{HPhi,Kawamura2017180}.

\appendix

\section{Zero Modes Structure}
\label{sec:zeromodes}

\begin{figure}[tt]
  \includegraphics[width=1.0\columnwidth]{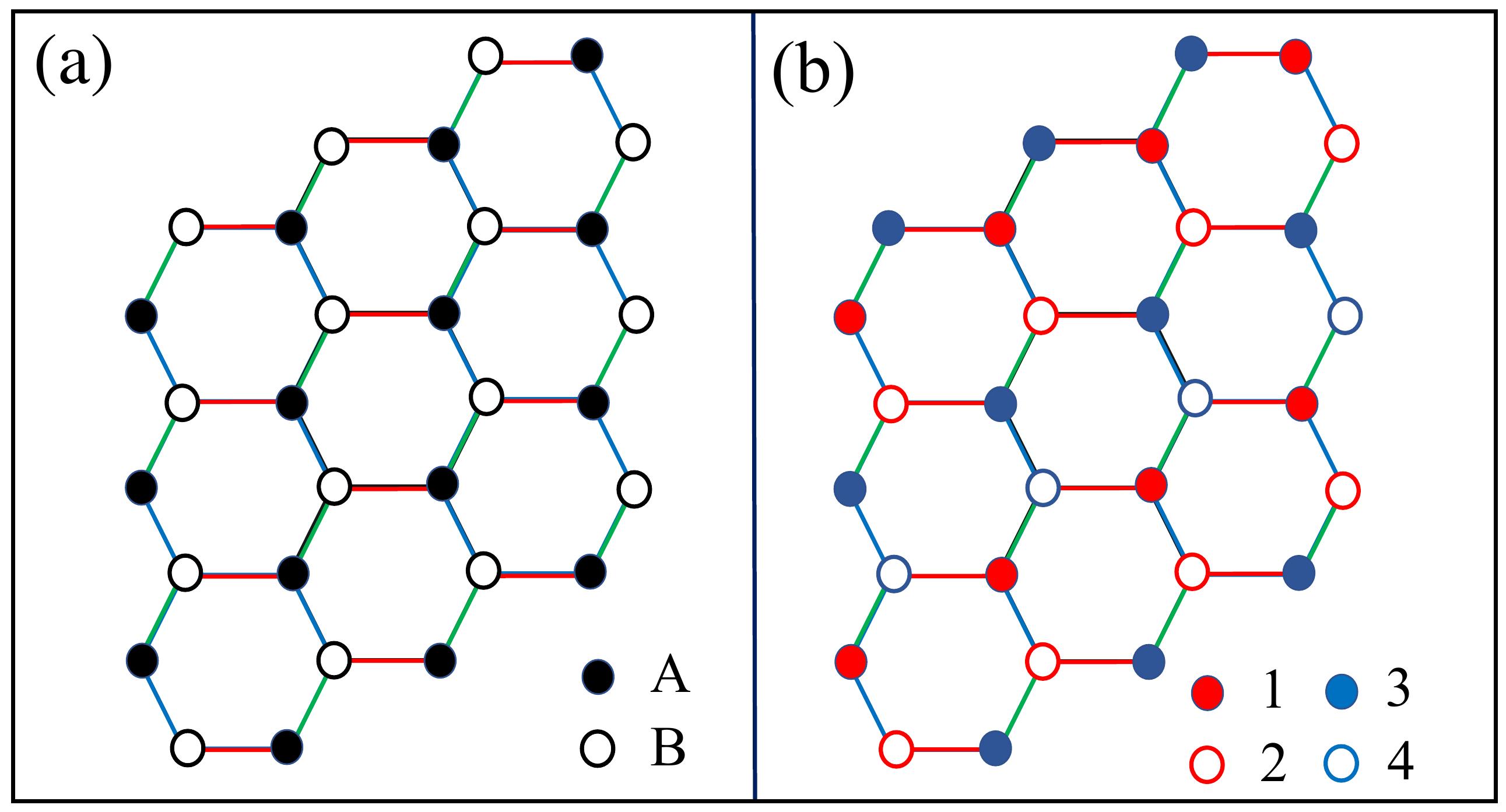}
  \caption{(a) Two-sublattice decomposition of the honeycomb
    lattice. (b) Four-sublattice decomposition of the honeycomb
    lattice.}
\label{subl}
\end{figure}

The momentum space distribution of the zero modes can be derived from
the set of constraints satisfied by the ground state manifold. As
reported by Rousochatzakis and
Perkins~\cite{rousochatzakis_classical_2017}, the classical ground
state of the AFM $\Gamma$ model satisfies the following constraints on
each bond of the lattice:
\begin{eqnarray}
  S^x_1 &=& -S^y_2, \;\;\; {\rm and} \;\;\;  S^y_1 = -S^x_2 \;\;\; {\rm for} \;\;\;{z-{\rm bonds}},
  \nonumber \\
  S^x_1 &=& -S^z_2, \;\;\; {\rm and} \;\;\; S^z_1 = -S^x_2 \;\;\; {\rm for} \;\;\;{y-{\rm bonds}},
  \nonumber \\
  S^y_1 &=& -S^z_2, \;\;\; {\rm and} \;\;\;  S^z_1 = -S^y_2 \;\;\; {\rm for} \;\;\;{x-{\rm bonds}},
  \label{gscond}
\end{eqnarray}
where $1$ and $2$ denote the two sites on the given
bond. Eq.~\ref{gscond} implies that the following identities hold for
any ground state configuration:
\begin{eqnarray}
  S^x_A &=& -S^y_B, \;\;\;   S^y_A= -S^x_B, \;\;\; S^x_A = -S^z_B, 
  \nonumber \\
  S^z_A &=& -S^x_B, \;\;\; S^y_A = -S^z_B, \;\;\;   S^z_A = -S^y_B,
  \label{gscond}
\end{eqnarray}
where $S^{\mu}_{A}= \sum_{j \in A} S^{\mu}_j$ and $S^{\mu}_{B}=
\sum_{j \in B} S^{\mu}_j$ and ($A$, $B$) denote the two sublattices of
the honeycomb lattice shown in Fig.~\ref{subl}~(a). The conditions
\eqref{gscond} lead to $S^x_A=S^y_A= S^z_A$, $S^x_B=S^y_B= S^z_B$ and
$S^{\mu}_A + S^{\mu}_B=0$, {\it implying that the ground states of the
  classical AFM $\Gamma$-model have no zero momentum component}. This
simple analysis proves the absence of elastic ($\omega=0$) spectral
weight at the ${\boldsymbol \Gamma}$ point.

Our next goal is to demonstrate the absence of elastic weight at the
${\bm X}$ points.  For this purpose, we divide the honeycomb lattice
into four sublattices, as shown in Fig.~\ref{subl}~(b). Without loss
of generality, we prove the statement for one of the three ${\bm X}$
points, say ${\bm X}_1$. The C$_6$ symmetry of the honeycomb lattice
guarantees that the result is the same for ${\bm X}_2$ and ${\bm
  X}_3$. The ${\bm X}_1$ component of a given spin configuration is:
\begin{equation}
  S^{\mu}_{{\bm X}_1} = \frac{2}{\sqrt{N}} \left [ S^{\mu}_1 +  S^{\mu}_2 - S^{\mu}_3 - S^{\mu}_4   \right ]
\end{equation}
where $S^{\mu}_{r} = \sum_{ j \in r} S^{\mu}_j $ and the integer index
$1\leq r \leq 4$ denotes each of the six sublattices and $N$ is the
total number of sites. From the general ground state condition
\eqref{gscond}, we obtain:
\begin{eqnarray}
  S^y_1 &=& -S^z_2, \;\;\;   S^z_1= -S^y_2, \;\;\; S^x_1 = -S^y_4, 
  \nonumber \\
  S^y_1 &=&  -S^x_4,  \;\;\; S^z_1 = -S^x_4, \;\;\; S^x_1 = -S^z_4,
  \nonumber \\
  S^x_2 &=& -S^y_3, \;\;\;   S^y_2= -S^x_3, \;\;\; S^x_2 = -S^z_3, 
  \nonumber \\
  S^z_2 &=& -S^x_3, \;\;\; S^y_3 = -S^z_4, \;\;\;   S^z_3= -S^y_4.
  \label{sixsub}
\end{eqnarray}
These identities give $S^x_1= -S^y_4 = S^z_3= S^x_2$, $S^x_3= -S^y_2 =
S^z_1= -S^x_4$, implying that $S^{x}_{{\bm X}_1} = 0 $. Similarly,
Eq.~\ref{sixsub} leads to $S^y_1= -S^z_2 = S^x_3=- S^y_2$, $S^z_1=
-S^x_4 = S^y_1=- S^z_2$, $S^z_3= -S^y_4 = S^x_1=- S^z_4$, implying
that $S^{y}_{{\bm X}_1} = S^{z}_{{\bm X}_1} = 0$.  In other words, the
ground state configurations of the AFM $\Gamma$ model have no ${\bm
  \Gamma}$ or ${\bm X}_{\mu}$ ($\mu=1,2,3$) components, implying the
absence of elastic ($\omega=0$) spectral weight at any of those
wave vectors.

\section{Symmetry analysis of \texorpdfstring{$S^{\alpha \beta}(\bm{r}, \bm{r'}, \omega)$}{$S(r, r', \omega)$}}
\label{sec:symm}

 Here we derive selection rules of the real space dynamical spin
structure factor of ${\cal H}$ for {\it arbitrary} spin $S$. For this
purpose, we introduce the six sublattice decomposition of the
honeycomb lattice that is depicted in
Fig.~\ref{symm}~\cite{rousochatzakis_classical_2017}. We 
demonstrate that six out of the nine components of the real space and real time magnetic structure factor,
\begin{equation}
  S^{\alpha \beta}({\bm r} , {\bm r}', t) = \langle S^{\alpha}_{\bm r} (0) S^{\beta}_{{\bm r}'} (t) \rangle,
\label{corr}
\end{equation}
always vanish as a consequence of the Hamiltonian symmetries that we discuss next.

It was noticed in Ref.~\onlinecite{rousochatzakis_classical_2017} that the Gamma model
\eqref{gammamodel}
is invariant under a set of three spin transformations acting on   each of the six sublattices depicted in Fig.~\ref{symm}.
\begin{itemize}
\item {If we decompose the full lattice into the white hexagons shown
    in Fig.~\ref{symm}, the Hamiltonian ${\cal H}$ is invariant under
    the symmetry operation:}
\begin{equation}
{\cal R}_a = 
\prod_{i \in \{ 1, 4\}} C_{2x}(i)   
\prod_{i' \in \{ 2, 5 \}} C_{2y}(i')  
\prod_{i'' \in \{ 3, 6 \}} C_{2z}(i''). 
\end{equation}
\item
{Similarly, a  lattice decomposition into the dark grey hexagons shown in Fig.~\ref{symm} reveals the symmetry operation:}
\begin{equation}
{\cal R}_b = 
\prod_{i \in \{ 6, 5\}} C_{2x}(i)   
\prod_{i' \in \{ 3, 4 \}} C_{2y}(i')  
\prod_{i'' \in \{ 1, 2 \}} C_{2z}(i''). 
\end{equation}

\item {Finally, a decomposition into the light grey hexagons in
    Fig.~\ref{symm} leads to the symmetry operation:}
\begin{equation}
  {\cal R}_c = 
  \prod_{i \in \{ 2, 3\}} C_{2x}(i)   
  \prod_{i' \in \{ 1, 6 \}} C_{2y}(i')  
  \prod_{i'' \in \{ 4, 5 \}} C_{2z}(i''). 
\end{equation}
\end{itemize}

We will derive now selection rules based on these symmetries. Given that these selection rules are excatly  the same
for any pair of sites ${\bm r}$ and ${\bm r}'$ belonging to  a given pair of sublattices $\nu$ and $\nu'$, we will use the  notation $\langle S^{\alpha}_{\nu} (0) S^{\beta}_{\nu'} (t) \rangle$ instead of $\langle S^{\alpha}_{\bm r} (0) S^{\beta}_{{\bm r}'} (t) \rangle$. We start by considering spin-spin correlators between sites {\it on the same sublattice}, i.e., both ${\bm r}$ and
${\bm r}'$ belong to the same sublattice $\nu$ ($1 \leq \nu \leq 6$). Off-diagonal contributions involve a product of two different spin components $S^{\alpha}_{\nu}$ and $S^{\beta}_{\nu}$ with $\alpha \neq \beta$. Because both spin operators belong to the same sublattice, they rotate about the same axis under the transformations  ${\cal R}_a $, ${\cal
  R}_b$ or ${\cal R}_c$. We can always choose the transformation ${\cal R}^{\;}_{\eta}$ that corresponds to a $\pi$ rotation about the $\alpha$-axis. Given that $\alpha$ and $\beta$ are different components, we have:
\begin{equation} 
  {\cal R}^{\dagger}_{\eta} S^{\alpha}_{\nu} {\cal R}^{\;}_{\eta} =  S^{\alpha}_{\nu},
  \;\;\; {\rm and} \;\;\;
  {\cal R}^{\dagger}_{\eta} S^{\beta}_{\nu} {\cal R}^{\;}_{\eta} = - S^{\beta}_{\nu},
\label{cond0}
\end{equation} 
implying that
\begin{eqnarray}
\langle S^{\alpha}_{\nu} (0) S^{\beta}_{\nu} (t) \rangle &=& {\rm Tr} [e^{- ({\cal H}/k_B T)} S^{\alpha}_{\nu} (0) S^{\beta}_{\nu} (t)] 
\nonumber \\
&=& {\rm Tr} [{\cal R}^{\dagger}_{\eta} e^{- ({\cal H}/k_B T)} S^{\alpha}_{\nu} (0) S^{\beta}_{\nu} (t){\cal R}^{\;}_{\eta}] 
\nonumber \\
&=& {\rm Tr} [ e^{- ({\cal H}/k_B T)} {\cal R}^{\dagger}_{\eta} S^{\alpha}_{\nu} (0) {\cal R}^{\;}_{\eta} {\cal R}^{\dagger}_{\eta}S^{\beta}_{\nu} (t){\cal R}^{\;}_{\eta}] 
\nonumber \\
&=& - \langle S^{\alpha}_{\nu} (0) S^{\beta}_{\nu} (t) \rangle =0.
\label{demo}
\end{eqnarray}
By using this result and and from the Hamiltonian symmetry under the product of a spin rotation by $2\pi/3$ about the $[111]$ direction and an orbital rotation by the same angle along the direction perpendicular to the plane of the honeycomb lattice, we obtain:
\begin{eqnarray}
\langle S^{\alpha}_{\nu} (0) S^{\beta}_{\nu} (t) \rangle = \delta_{\alpha \beta}
\langle S^{z}_{\nu} (0) S^{z}_{\nu} (t) \rangle
\end{eqnarray}
for general values of $\alpha$ and $\beta$.

We consider now the spin-spin correlator \eqref{corr} for ${\bm r}$ and ${\bm
  r}'$ belonging to {\it different} sublattices with the {\it same
  parity} ($\nu \neq \nu'$ and $\nu + \nu'$ even). For any diagonal component ($\mu =\nu$), it is easy to verify
that at least one of the three symmetry transformations, ${\cal R}_a $, ${\cal
  R}_b$ or ${\cal R}_c$ changes the sign of only  one of the two spin operators:
\begin{equation} 
  {\cal R}^{\dagger}_{\eta} S^{\alpha}_{\nu} {\cal R}^{\;}_{\eta} = \pm S^{\alpha}_{\nu},
  \;\;\; {\rm and} \;\;\;
  {\cal R}^{\dagger}_{\eta} S^{\alpha}_{\nu'} {\cal R}^{\;}_{\eta} = \mp S^{\alpha}_{\nu'}.
\label{cond}
\end{equation}
Here $\eta=a, b$ or $c$ denotes the transformation that satisfies
\eqref{cond}.  Note that ${\cal R}^{\dagger}_{\eta}= {\cal
  R}^{\;}_{\eta}$. Once again, following the same procedure as in Eq.~\eqref{demo}, we obtain
\begin{eqnarray}
\langle S^{\alpha}_{\nu} (0) S^{\alpha}_{\nu'} (t) \rangle  =0.
\end{eqnarray}
By using a similar procedure, we can demonstrate that three, out of the six, off-diagonal correlators between different sublattices with the same parity are also equal to zero: 
\begin{eqnarray}
\langle S^{x}_{1} (0) S^{y}_{3} (t) \rangle &=& \langle S^{y}_{1} (0) S^{z}_{3} (t) \rangle =
\langle S^{z}_{1} (0) S^{x}_{3} (t) \rangle =0,
\nonumber \\
\langle S^{x}_{1} (0) S^{z}_{5} (t) \rangle &=& \langle S^{z}_{1} (0) S^{y}_{5} (t) \rangle =
\langle S^{y}_{1} (0) S^{x}_{5} (t) \rangle =0,
\nonumber \\
 \langle S^{y}_{2} (0) S^{z}_{4} (t) \rangle &=& \langle S^{z}_{2} (0) S^{x}_{4} (t) \rangle =
\langle S^{x}_{2} (0) S^{y}_{4} (t) \rangle =0,
\nonumber \\
\langle S^{y}_{2} (0) S^{x}_{6} (t) \rangle &=& \langle S^{x}_{2} (0) S^{z}_{6} (t) \rangle =
\langle S^{z}_{2} (0) S^{y}_{6} (t) \rangle =0,
\nonumber \\
\end{eqnarray}
We note that  $\langle S^{\alpha}_{\bm r} (0) S^{\beta}_{{\bm r}'} (t) \rangle=0$ implies 
$\langle S^{\beta}_{{\bm r}'} (0) S^{\alpha}_{\bm r} (t) \rangle=0$.

Similarly, using the symmetries ${\cal R}_a $, ${\cal R}_b$  and ${\cal R}_c$, we can demonstrate that:
\begin{eqnarray}
\langle S^{x}_{1} (0) S^{x}_{6} (t) \rangle &=& \langle S^{z}_{1} (0) S^{z}_{6} (t) \rangle =
\langle S^{x}_{1} (0) S^{y}_{6} (t) \rangle 
\nonumber \\
&=& \langle S^{y}_{1} (0) S^{x}_{6} (t) \rangle = \langle S^{z}_{1} (0) S^{y}_{6} (t) \rangle 
\nonumber \\
&=& \langle S^{y}_{1} (0) S^{z}_{6} (t) \rangle =0,
\nonumber \\
\langle S^{x}_{1} (0) S^{x}_{2} (t) \rangle &=& \langle S^{y}_{1} (0) S^{y}_{2} (t) \rangle =
\langle S^{x}_{1} (0) S^{z}_{2} (t) \rangle 
\nonumber \\
&=& \langle S^{z}_{1} (0) S^{x}_{2} (t) \rangle = \langle S^{z}_{1} (0) S^{y}_{2} (t) \rangle 
\nonumber \\
&=& \langle S^{y}_{1} (0) S^{z}_{2} (t) \rangle =0,
\nonumber \\
\langle S^{y}_{1} (0) S^{y}_{4} (t) \rangle &=& \langle S^{z}_{1} (0) S^{z}_{4} (t) \rangle =
\langle S^{x}_{1} (0) S^{z}_{4} (t) \rangle  
\nonumber \\
&=& \langle S^{z}_{1} (0) S^{x}_{4} (t) \rangle = \langle S^{x}_{1} (0) S^{y}_{4} (t) \rangle 
\nonumber \\
&=& \langle S^{y}_{1} (0) S^{x}_{4} (t) \rangle =0,
\nonumber \\
\langle S^{x}_{2} (0) S^{x}_{1} (t) \rangle &=& \langle S^{y}_{2} (0) S^{y}_{1} (t) \rangle =
\langle S^{x}_{2} (0) S^{z}_{1} (t) \rangle 
\nonumber \\
&=& \langle S^{z}_{2} (0) S^{x}_{1} (t) \rangle  = \langle S^{z}_{2} (0) S^{y}_{1} (t) \rangle 
\nonumber \\
&=& \langle S^{y}_{2} (0) S^{z}_{1} (t) \rangle =0,
\nonumber \\
\langle S^{y}_{2} (0) S^{y}_{3} (t) \rangle &=& \langle S^{z}_{2} (0) S^{z}_{3} (t) \rangle =
\langle S^{x}_{2} (0) S^{z}_{3} (t) \rangle 
\nonumber \\
&=& \langle S^{z}_{2} (0) S^{x}_{3} (t) \rangle  = \langle S^{x}_{2} (0) S^{y}_{3} (t) \rangle 
\nonumber \\
&=& \langle S^{y}_{2} (0) S^{x}_{3} (t) \rangle =0,
\nonumber \\
\langle S^{x}_{2} (0) S^{x}_{5} (t) \rangle &=& \langle S^{z}_{2} (0) S^{z}_{5} (t) \rangle =
\langle S^{x}_{2} (0) S^{y}_{5} (t) \rangle 
\nonumber \\
&=& \langle S^{y}_{2} (0) S^{x}_{5} (t) \rangle  = \langle S^{z}_{2} (0) S^{y}_{5} (t) \rangle 
\nonumber \\
&=& \langle S^{y}_{2} (0) S^{z}_{5} (t) \rangle =0, 
\nonumber \\
&=& \langle S^{z}_{6} (0) S^{x}_{5} (t) \rangle = \langle S^{x}_{6} (0) S^{y}_{5} (t) \rangle 
\nonumber \\
&=& \langle S^{y}_{6} (0) S^{x}_{5} (t) \rangle =0.
\end{eqnarray}
It is clear then that the symmetries ${\cal R}_a $, ${\cal R}_b$ and ${\cal R}_c$
constrain six components of the real space spin structure factor to
be identically zero. The six components that vanish depend on the two sublattices to which the 
vectors ${\bm r}$ and ${\bm r}'$ belong to.

\section{MSR treatment of the classical Kitaev model}
\label{sec:msrkitaev}

In the honeycomb Kitaev model, defined by
\begin{equation}
  \label{eq:Kitaev}
  K_{ij}^{\alpha\beta}=\left\{
    \begin{array}{cc}
      K & \alpha=\beta=\braket{ij} \\ 0 & {\rm otherwise}
    \end{array}\right. ,
\end{equation}
each component of a given spin is correlated only with one component
of one neighboring spin -- the one it interacts with. Thus,
\begin{equation}
  \label{eq:G0Kitaev}
  \left(G_0^{-1}\right)_{ij}^{\alpha\beta} = \left\{
    \begin{array}{cc}
      -i\omega+\gamma\Delta & \alpha=\beta,i=j 
      \\ \gamma K & \alpha=\beta=\braket{ij} 
      \\ 0 & {\rm otherwise}
    \end{array}\right.
\end{equation}
from which we find that $C_0$ becomes a $2\times 2$ matrix, given by
\begin{equation}
  \label{eq:G0K2} C_0(\omega) = \frac{1}{2}\sum_{m=\pm}\left(
    \begin{array}{cc} 1 & m \\ m & 1
    \end{array}\right)\frac{1}{\omega^2+\gamma^2(\Delta+mK)^2}.  
\end{equation}
Physically, this indicates that correlations decay at two
characteristic rates, as given by the two eigenvalues
$\gamma(\Delta\pm K)$.  The dynamic structure factor is obtained by
taking the Fourier transform, $S({\bf
  q},\omega)=3C_{11}(\omega)+\sum_{\bm{\delta}}e^{i{\bf
    q}\cdot\bm{\delta}} C_{12}(\omega)$, where $\bm{\delta}$ denotes
the vectors connecting nearest neighbors. Evidently, for the
anti-ferromagnetic Kitaev model, correlations at ${\bf q}=0$ decay
only at the fast rate, $\gamma(\Delta+K)$, leading to a depletion in
the dynamic structure factor at low frequencies.

\begin{figure}[b]
  \centering
  \includegraphics[width=\linewidth]{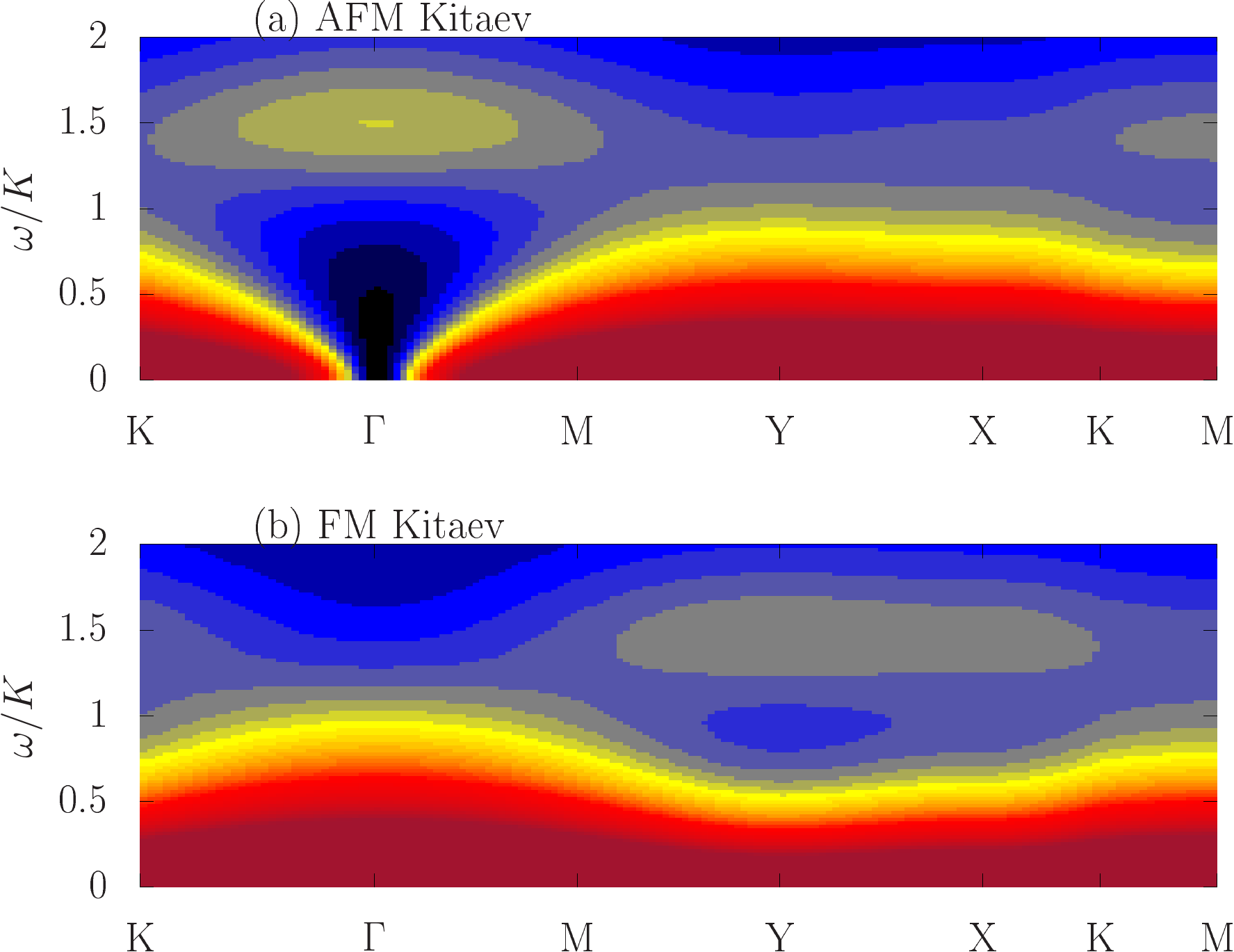} 
  \caption{Dynamic structure factor as obtained in Eq. (\ref{eq:CK2}),
    for (a) the AFM Kitaev $K>0$, and (b) the FM Kitaev $K<0$
    models. Here we used $\gamma=0.25$, $\Delta=1.05K$ and
    $g^2T=0.1K$.}
  \label{fig:msr_kitaev}
\end{figure}

At finite $g>0$, the infinite series in
  Eq. (\ref{eq:C}), becomes a $2\times 2$ matrix equation, with
\begin{equation}
  \label{eq:G0K2} G_0(\omega) = \left(
    \begin{array}{cc} -i\omega+\gamma\Delta & \gamma K \\ \gamma K &
-i\omega+\gamma\Delta
    \end{array}\right)^{-1},
\end{equation} 
while the self energy calculation yields 
\begin{equation}
  \label{eq:SEK2}
\Sigma(\omega)=-g^2K^2\frac{1}{-i\omega+2\gamma\Delta}\frac{T}{\Delta^2-K^2}
\left( \begin{array}{cc} \Delta & K \\ K & \Delta
    \end{array}\right).
\end{equation} Using Eq. (\ref{eq:C}), we obtain
\begin{eqnarray}
  \label{eq:CK2} C(\omega)&\approx&\frac{1}{2}\sum_{m=\pm 1}
  \left(\begin{array}{cc} 1 & m \\ m & 1
    \end{array}\right)  \\ & & \quad\quad \times\frac{2\gamma 
    T}{\left|-i\omega+\gamma(\Delta+mK)+\frac{g^2K^2T}{\Delta^2-K^2}
      \frac{\Delta+mK}{-i\omega+2\gamma\Delta}\right|^2}.\nonumber
\end{eqnarray} 
We obtain the dynamic structure factor by Fourier transforming this
result, see Fig. \ref{fig:msr_kitaev}. Note, for example, that only
the $m=1$ term contributes to the dynamic structure factor at ${\bf
  q}=0$, which is suppressed at low frequencies for the AFM Kitaev,
$K>0$. Evidently, the self-energy, Eq. (\ref{eq:SEK2}), is larger for
$m=1$, producing a precession peak at finite $\omega$, where the low
frequency correlations are suppressed.

\section{Details of the ED calculation}
\label{details_of_ED}
When every eigenvalue $\{E_{\nu} \}$ and eigenvector $\{\ket{\nu}\}$ of the hamiltonian $\mathcal{H}$
are known, the Green's function at a finite temperature $\beta^{-1}$ is given as 
\eqsa{
\displaystyle 
\mathcal{G}^{AB}_{\beta}(\omega)
=
\sum_{\nu,\mu}
\frac{e^{-\beta E_{\nu}}}{Z(\beta)}
\frac{\bra{\nu}
{A}^{\dagger}
\ket{\mu}
\bra{\mu}
{B}^{\ }
\ket{\nu}}
{\omega+i\eta+E_{\nu}-E_{\mu}},
}
where $Z$ is the partition function of the system defined as $\displaystyle Z(\beta)=\sum_{\nu}e^{-\beta E_{\nu}}$.
For later use, we rewrite the above expression of $\mathcal{G}^{AB}_{\beta}$ as
\eqsa{
\mathcal{G}^{AB}_{\beta}(\omega)
=
\displaystyle 
\sum_{\nu}
\frac{e^{-\beta E_{\nu}}}{Z(\beta)}
\bra{\nu}
{A}^{\dagger}
\frac{1}{\omega+i\eta+E_{\nu}-\mathcal{H}}
{B}^{\ }
\ket{\nu}.
\label{chiAB}
\nonumber\\
}

Here, we reformulate Eq.(\ref{chiAB}) with a typical pure state~\cite{Imada_Takahashi,skilling2013maximum,
PhysRevB.47.7929,
PhysRevB.49.5065,
PhysRevE.62.4365,PhysRevLett.108.240401,PhysRevLett.111.010401} $\ket{\psi_{\beta}}$
to avoid using the whole set of $E_{\nu}$ and $\ket{\nu}$.
First, we note that the normalized typical state is naively expected to behave as
\eqsa{
\ket{\psi_{\beta}}
\sim \sum_{\nu}e^{i\varphi_{\nu}}\frac{e^{-\frac{\beta}{2} E_{\nu}}}{\sqrt{Z(\beta)}}\ket{\nu},
}
where $\varphi_{\nu} \in [0,2\pi)$ are random numbers. 
By introducing a projection operator,
\eqsa{
\hat{P}_{\nu} = \ket{\nu}\bra{\nu},
}
we
rewrite the formula based on canonical ensemble, Eq.(\ref{chiAB}), as
\eqsa{
\mathcal{G}^{AB}_{\beta}(\zeta)
\sim
\sum_{\nu}
\bra{\psi_{\beta}}
\hat{P}_{\nu}
{A}^{\dagger}
\frac{1}{\zeta+E_{\nu}-\mathcal{H}}
{B}
\hat{P}_{\nu}
\ket{\psi_{\beta}}.
}
Thus far,
the exact projection operator $\hat{P}_{\nu}$ requires the whole set of $\ket{\nu}$. 

The important step is to find an efficient implementation of the projection operator $\hat{P}_{\nu}$. 
Although there is no $\mathcal{O}(N_{\rm F})$ implementation of the exact $\hat{P}_{\nu}$ in the literature as far as we know,
where $N_{\rm F}$ is the dimension of the Fock space,
there is a filter operator~\cite{doi:10.1143/ptp/4.4.514,Sakurai2003119,ikegami2010filter,Shimizu201613}
that constructs equi-energy shells and is realizable with the numerical cost of $\mathcal{O}(N_{\rm F})$
by employing the shifted Krylov method~\cite{frommer2003bicgstab}, as follows.

The filter operator~\cite{doi:10.1143/ptp/4.4.514} is defined by integrating the resolvent of $\hat{H}$ along a contour $C_{\gamma,\rho}$ defined
by $z=\rho e^{i\theta} + \gamma$ with $0\leq \theta < 2\pi $ as
\eqsa{
\hat{P}_{\gamma,\rho}=\frac{1}{2\pi i}\oint_{C_{\gamma,\rho}}\frac{dz}{z-\mathcal{H}}.
}
If the filter operator is applied to an arbitrary wave function $\ket{\phi}=\sum_{\nu} d_{\nu} \ket{\nu}$,
the operator filters the eigenvectors with the eigenvalues $E_{\nu} \not\in (\gamma-\rho,\gamma+\rho)$. 
When a small $\gamma$ limit is taken, the filter operator realizes a microcanonical ensemble.
The filter operator is practically implemented as a Reimann sum~\cite{Sakurai2003119,ikegami2010filter}:
The discretized filter operator is defined as
\eqsa{
\hat{P}_{\gamma,\rho,M}=
\frac{1}{M}
\sum_{j=1}^{M}
\frac{\rho e^{i\theta_j}}{\rho e^{i\theta_j}+\gamma -\mathcal{H}},
\label{dfilterP}
}
where $\theta_j = 2\pi (j-1/2)/M$.
Multiplication of $\hat{P}_{\gamma,\rho,M}$ to
a wave function is simply realized by the shifted Krylov subspace method
while it is hardly achievable by the standard Lanczos algorithm.

By introducing an appropriate energy grid measured from the low-energy onset $E_{\rm b}$ in energy axis,
\eqsa{
\mathcal{E}_m=E_{\rm b}+(2m+1)\epsilon,
}
the set of the filter operators $\{\hat{P}_{\mathcal{E}_m,\epsilon,M}\}$
with the discretization parameters,
\eqsa{\mbox{\boldmath$\delta$}=(E_{\rm b},\epsilon,M),\label{paramdis}}
indeed replace the projection operators $\{\hat{P}_n\}$.
The filtered typical state given by
\eqsa{
\ket{\phi_{\beta,\mbox{\boldmath$\delta$}}^{m}}
=
\hat{P}_{\mathcal{E}_m,\epsilon,M}
\ket{\phi_{\beta}}\label{filtered}
}
is a random vector residing in an equi-energy shell $(\mathcal{E}_m-\epsilon,\mathcal{E}_m+\epsilon)$,
which corresponds to a microcanonical ensemble.

A representation of the Green's function is thus achieved by employing the filtered typical pure states
$\{\ket{\psi_{\beta,\mbox{\boldmath$\delta$}}^{m}}\}$
as
\eqsa{
\widetilde{\mathcal{G}}_{\beta,\mbox{\boldmath$\delta$}}^{AB}(\zeta)
=
\sum_{m = 0}^{L-1}
\bra{\psi_{\beta,\mbox{\boldmath$\delta$}}^{m}}
{A}^{\dagger}
\frac{1}{\zeta+\mathcal{E}_m -\mathcal{H}}
{B}
\ket{\psi_{\beta,\mbox{\boldmath$\delta$}}^{m}}.\label{GAB_TPS}
}
After taking appropriate limits and average over the distribution of the initial random vectors of the typical pure states,
we indeed replace the canonical ensemble prescription by the typical pure state formula.
By setting $\zeta=\omega+i\eta$ and \eqsa{
  {A}={B}=\hat{S}_{+\mbox{\boldmath$Q$}}^{\alpha}\equiv
  N^{-1/2}\sum_{\ell}e^{+i\mbox{\boldmath$Q$}\cdot\mbox{\boldmath$R$}_{\ell}}{S}^{\alpha}_{\ell},
}
in Eq.(\ref{GAB_TPS}),
we obtain the dynamical spin structure factor at a momentum $\mbox{\boldmath$Q$}$ and a frequency $\omega$ as,
\eqsa{
\widetilde{S}_{\beta,\mbox{\boldmath$\delta$}}(\mbox{\boldmath$Q$},\omega)
&=&
-\frac{1}{\pi}
{\rm Im}
\sum_{\alpha=x,y,z}
\sum_{m=0}^{L-1}
\bra{\psi_{\beta,\mbox{\boldmath$\delta$}}^{m}}
{S}^{\alpha}_{-\mbox{\boldmath$Q$}}
\nonumber\\
&&\times
\frac{1}{\omega+i\eta+\mathcal{E}_m -\mathcal{H}}
{S}^{\alpha}_{+\mbox{\boldmath$Q$}}
\ket{\psi_{\beta,\mbox{\boldmath$\delta$}}^{m}},\label{SQomega_TPS}
}
where ${S}_{\ell}^{\alpha}$ ($\alpha=x,y,z$) is an $S$=1/2 spin operator.

In the present paper, we set $E_{\rm b}$ in Eq.(\ref{paramdis}) as $E_{\rm b} = -8.6$ $(< E_0 \simeq -8.57)$
for the 24 site cluster of the $\Gamma$ model.
The distance among the energy grid points $2\epsilon$ is set
as $(E_{\rm cut}-E_{\rm b})/L$, where $L=128$ and $E_{\rm cut}$ is chosen depending on $T$ as $E_{\rm cut} = \max\{\min\{|E_{\rm b}|,T\ln 10^{14} + E_{\rm b}\}, 4 \}$.
A random vector is chosen as a typical pure state $|\psi_{0}\rangle$ at infinite temperature.
Then, a typical pure state at finite inverse temperature $\beta$ is
given by $|\psi_{\beta}\rangle = e^{-\beta\mathcal{H}/2}|\psi_{0}\rangle/\langle \psi_{0}| 
e^{-\beta\mathcal{H}}|\psi_{0}\rangle$.

\section{High temperature expansion}
\label{sec:highT}

\begin{figure}[b]
\centering
\includegraphics[width=4.0cm]{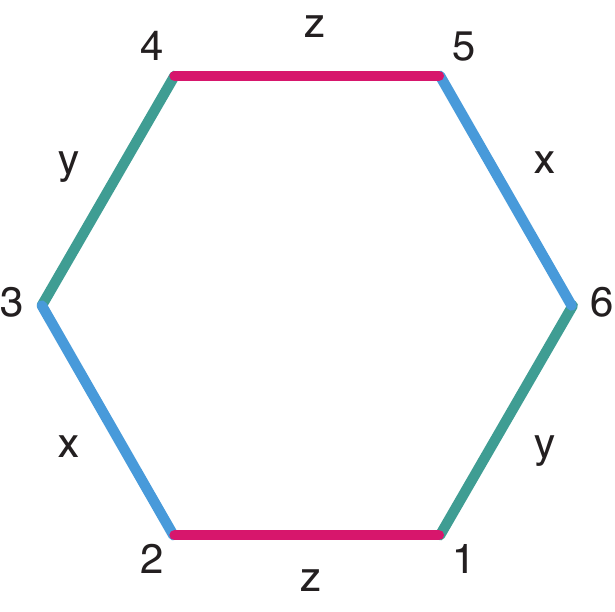}
\caption{(color online):
Site indices used in calculations of spin correlations.
}
\label{fig_hex}
\end{figure}

Here, high temperature expansion (small $\beta$ expansion) of static spin correlations
is obtained up to third order of $\beta$. 
Within the language of thermal pure quantum states~\cite{PhysRevLett.111.010401} one can write the
finite temperature expectation value of an operator ${O}$ as
\eqsa{
\langle {O}\rangle=
\frac{
\displaystyle
\mathbb{E}
\left[
\left(\sum_{\nu}c^{\ast}_{\nu}\bra{\nu}\right)
e^{-\beta\mathcal{H}/2}
{O}
e^{-\beta\mathcal{H}/2}
\left(\sum_{\mu}c_{\mu}\ket{\mu}\right)
\right]
}{
\displaystyle
\mathbb{E}
\left[
\left(\sum_{\nu}c^{\ast}_{\nu}\bra{\nu}\right)
e^{-\beta\mathcal{H}}
\left(\sum_{\mu}c_{\mu}\ket{\mu}\right)
\right]
},
\nonumber\\
}
where $\{c_{\nu}\}$ is a set of random complex numbers that satisfy
the normalization $\sum_{\nu}|c_{\nu}|^2 = 1$ and
the average $\mathbb{E}[\cdots]$ is taken over the probability distribution of
the random complex numbers.
The denominator is estimated as
\eqsa{
&&\mathbb{E}
\left[
\left(\sum_{\nu}c^{\ast}_{\nu}\bra{\nu}\right)
e^{-\beta\mathcal{H}}
\left(\sum_{\mu}c_{\mu}\ket{\mu}\right)
\right]
\nonumber\\
&&\displaystyle
=1+
\frac{\beta^2}{2N_{\rm F}}{\rm tr}[\mathcal{H}^2]+
\frac{\beta^4}{24N_{\rm F}}{\rm tr}[\mathcal{H}^4]+\mathcal{O}(\beta^6),
}
where $N_{\rm F}$ is the Fock space dimension and
$\mathbb{E}[c_{\nu}^{\ast}c_{\mu}]=\delta_{\nu,\mu}/N_{\rm F}$ is used.
Then, the numerator is estimated by expanding it with respect to $\beta$.
The first term is
\eqsa{
&&\mathbb{E}
\left[
\left(\sum_{\nu}c^{\ast}_{\nu}\bra{\nu}\right)
{O}
\left(\sum_{\mu}c_{\mu}\ket{\mu}\right)
\right]
\nonumber\\
&&=
\sum_{\nu,\mu}
\mathbb{E}[c^{\ast}_{\nu}c_{\mu}]\bra{\nu}{O}\ket{\mu}
\nonumber\\
&&=
\frac{1}{N_{\rm F}}\sum_{\nu}\bra{\nu}{O}\ket{\nu}
=\frac{1}{N_{\rm F}}{\rm tr}[{O}].
}
The higher order terms are given by
\eqsa{
({\rm 2nd\ term})&=&
-\frac{\beta}{2N_{\rm F}}{\rm tr}[{O}\mathcal{H}+\mathcal{H}{O}],
\label{app_2nd}
\\
({\rm 3rd\ term})&=&
\frac{\beta^2}{4N_{\rm F}}
{\rm tr}[\frac{1}{2}{O}\mathcal{H}^2+\mathcal{H}{O}\mathcal{H}+\frac{1}{2}\mathcal{H}^2{O}],
\label{app_3rd}
\\
({\rm 4th\ term})&=&
-\frac{\beta^3}{8N_{\rm F}}
{\rm tr}[\frac{1}{6}{O}\mathcal{H}^3+\frac{1}{2}\mathcal{H}{O}\mathcal{H}^2
\nonumber 
\\ & & \quad\quad +\frac{1}{2}\mathcal{H}^2{O}\mathcal{H}+\frac{1}{6}\mathcal{H}^3{O}].
\label{app_4th}
}

When ${O}$ is a spin-spin correlation defined by $\prod_{\ell} \left(S^{x}_{\ell}\right)^{n_{\ell x}}\left(S^{z}_{\ell}\right)^{n_{\ell y}}\left(S^{z}_{\ell}\right)^{n_{\ell z}}$
 ($n_{\ell \alpha}=0,1$),
${\rm tr}[{O}]=0$ or, at least, ${\rm tr}[{O}]\ll N_{\rm F}$ for $\sum_{\ell,\alpha}n_{\ell \alpha}\neq 0$,
because ${\rm tr}[{O}]/N_{\rm F}$ is the expectation value $\langle {O} \rangle$ at $\beta=0$.
Only if ${O}$ is the identity matrix, ${\rm tr}[{O}]=N_{\rm F}$.

\subsection{$yx$ correlation for nearest neighbor $z$ bond}
The high temperature expansion of $\langle S^y_{\mbox{\boldmath$0$}}S^x_{\mbox{\boldmath$r$}}\rangle$ for the nearest neighbor $z$ bonds is given as follows.
The lowest order of a finite ${\rm tr}[\mathcal{H}^m {S}^y_2 {S}^x_1 \mathcal{H}^n]/N_{\rm F}$ (see Fig.~\ref{fig_hex} for the site indices)
is given by
\eqsa{
\Gamma {S}^y_2 {S}^x_1 \cdot {S}^y_2 {S}^x_1 = \frac{\Gamma}{16}.
}
When we set $O={S}^y_2 {S}^x_1$ in Eq. (\ref{app_2nd}),
\eqsa{
({\rm 2nd\ term})&=& -\frac{\beta\Gamma}{16}.
}

\subsection{$zz$ correlation for nearest neighbor $z$ bond}
The high temperature expansion of $\langle S^z_{\mbox{\boldmath$0$}}S^z_{\mbox{\boldmath$r$}}\rangle$
for the nearest neighbor $z$ bonds is given as follows.
The lowest order of a finite ${\rm tr}[\mathcal{H}^m {S}^z_2 {S}^z_1 \mathcal{H}^n]/N_{\rm F}$ (see Fig.~\ref{fig_hex} for the site indices)
is, for example, given by
\eqsa{
\Gamma^2 {S}^z_2 {S}^z_1 {S}^x_2 {S}^y_1 {S}^y_2 {S}^x_1 = \frac{\Gamma^2}{64}.
}
Then, if we set ${O}={S}^z_2 {S}^z_1$ in Eqs. (\ref{app_2nd}) and (\ref{app_3rd}),
\eqsa{
({\rm 2nd\ term})&=& 0,
\\
({\rm 3rd\ term})&=&
\frac{\beta^2\Gamma^2}{64}.
}
\subsection{$zz$ correlation for third nearest neighbor}
The high temperature expansion
of $\langle S^z_{\mbox{\boldmath$0$}}S^z_{\mbox{\boldmath$r$}}\rangle$
for the third nearest neighbor pairs across hexagons is given as follows.
The lowest order of a finite ${\rm tr}[\mathcal{H}^m {S}^z_3 {S}^z_6 \mathcal{H}^n]/N_{\rm F}$ (see Fig.~\ref{fig_hex} for the site indices)
is, for example, given by
\eqsa{
\Gamma^3 {S}^z_3 {S}^z_6 \cdot {S}^z_3 {S}^y_2 \cdot {S}^y_2 {S}^x_1 
\cdot {S}^z_1 {S}^z_6
= \frac{\Gamma^3}{256},
}
or
\eqsa{
\Gamma^3 {S}^z_3 {S}^z_6 \cdot {S}^z_3 {S}^x_4 \cdot {S}^x_4 {S}^y_5 
\cdot {S}^y_5 {S}^z_6
= \frac{\Gamma^3}{256}.
}
Then, if we set ${O}={S}^z_3 {S}^z_6$ in Eqs. (\ref{app_2nd}), (\ref{app_3rd}), and (\ref{app_4th}),
\eqsa{
({\rm 2nd\ term})&=& 0,
\\
({\rm 3rd\ term})&=& 0,
\\
({\rm 4th\ term})&=& \frac{3\Gamma^3 \beta^3}{384}.
}

\bibliography{kgamma} 

\end{document}